\documentclass[12pt]{article}
\usepackage{fullpage,cite,epsfig,psfrag,graphics,amsbsy,amssymb,amsmath,wasysym,subcaption}
\usepackage[T1]{fontenc}
\usepackage{hyperref}
\usepackage{caption}
\newcommand{\ba}{\begin{aligned}}
\newcommand{\ea}{\end{aligned}}
\newcommand{\ben}{\begin{displaymath}}
\newcommand{\een}{\end{displaymath}}
\newcommand{\bean}{\begin{eqnarray*}}
\newcommand{\eean}{\end{eqnarray*}}
\newcommand{\beq}{\begin{equation}}
\newcommand{\eeq}{\end{equation}}
\newcommand{\bea}{\begin{eqnarray}}
\newcommand{\eea}{\end{eqnarray}}

\newcommand{\be}{\begin{equation}}
\newcommand{\ee}{\end{equation}}
\newcommand{\bq}{\begin{eqnarray}}
\newcommand{\eq}{\end{eqnarray}}

\def\math{\mathsurround=0pt }
\def\leftrightarrowfill{$\math \mathord\leftarrow \mkern-6mu
 \cleaders\hbox{$\mkern-2mu \mathord- \mkern-2mu$}\hfill
 \mkern-6mu \mathord\rightarrow$}
\def\overleftrightarrow#1{\vbox{\ialign{##\crcr
     \leftrightarrowfill\crcr\noalign{\kern-1pt\nointerlineskip}
     $\hfil\displaystyle{#1}\hfil$\crcr}}}

\let\l=\lambda

 \def\bd{\begin{document}} \def\ed{\end{document}}
\def\ds{\documentstyle} \let\fr=\frac \let\bl=\bigl \let\br=\bigr
\let\Br=\Bigr \let\Bl=\Bigl
\let\bm=\bibitem
\let\na=\nabla
\let\pa=\partial \let\ov=\overline
\def\ft#1#2{{\textstyle{{\scriptstyle #1}\over {\scriptstyle #2}}}}
\def\fft#1#2{{#1 \over #2}}
\def\vp{\varphi}
\def\sst#1{{\scriptscriptstyle #1}}
\def\oneone{\rlap 1\mkern4mu{\rm l}}
\def\td{\tilde}
\def\wtd{\widetilde}
\def\dalemb#1#2{{\vbox{\hrule height .#2pt
        \hbox{\vrule width.#2pt height#1pt \kern#1pt
                \vrule width.#2pt}
        \hrule height.#2pt}}}
\def\square{\mathord{\dalemb{6.8}{7}\hbox{\hskip1pt}}}
\def\wtd{\widetilde}
\def\R{\rlap{\rm I}\mkern3mu{\rm R}}
\def\im{{\rm i}}
\def\tilg{\tilde{g}}
\def\tilF{\tilde{F}}
\def\tilA{\tilde{A}}
\def\varf{\varphi}
\def\tilf{\tilde{\phi}}
\def\tilh{\tilde{h}}
\def\rme{{\rm e}}
\def\ep{\epsilon}
\def\0{{(0)}}
\def\9{{(9)}}
\def\8{{(8)}}
\def\7{{(7)}}
\def\6{{(6)}}
\def\5{{(5)}}
\def\4{{(4)}}
\def\3{{(3)}}
\def\2{{(2)}}
\def\1{{(1)}}
\newcommand{\trace}{{\rm Tr}}
\newcommand{\ub}{\overline{U}}
\newcommand{\vb}{\overline{V}}
\newcommand{\uh}{\widehat{U}}
\newcommand{\vh}{\widehat{V}}
\newcommand{\ubh}{\overline{\widehat{U}}}
\newcommand{\vbh}{\overline{\widehat{V}}}
\newcommand{\lb}{\bar{\l}}
\newcommand{\Fb}{\overline{F}}
\newcommand{\Fh}{\widehat{F}}
\newcommand{\Fbh}{\overline{\widehat{F}}}
\newcommand{\Ab}{\overline{A}}
\newcommand{\Ah}{\widehat{A}}
\newcommand{\Abh}{\overline{\widehat{A}}}
\newcommand{\Gb}{\overline{G}}
\newcommand{\Gh}{\widehat{G}}
\newcommand{\Gbh}{\overline{\widehat{G}}}
\newcommand{\Pb}{\overline{P}}
\newcommand{\Ph}{\widehat{P}}
\newcommand{\Pbh}{\overline{\widehat{P}}}
\newcommand{\Qb}{\overline{Q}}
\newcommand{\Qh}{\widehat{Q}}
\newcommand{\Qbh}{\overline{\widehat{Q}}}
\newcommand{\Bb}{\overline{B}}
\newcommand{\Bh}{\widehat{B}}
\newcommand{\Bbh}{\overline{\widehat{B}}}
\newcommand{\fhns}{\hat{F}^{\rm (NS)}}
\newcommand{\fhrr}{\hat{F}^{\rm (RR)}}
\newcommand{\ahns}{\hat{A}^{\rm (NS)}}
\newcommand{\ahrr}{\hat{A}^{\rm (RR)}}
\newcommand{\hhrr}{\hat{H}^{\rm (RR)}}
\newcommand{\hchi}{\hat{\chi}}
\newcommand{\hphi}{\hat{\phi}}
\newcommand{\htau}{\hat{\tau}}
\newcommand{\cG}{{\cal G}}
\newcommand{\cGb}{\overline{{\cal G}}}
\newcommand{\cH}{{\cal H}}
\newcommand{\cP}{{\cal P}}
\newcommand{\cPb}{\overline{{\cal P}}}
\newcommand{\cQ}{{\cal Q}}
\newcommand{\cQb}{\overline{{\cal Q}}}
\newcommand{\cM}{{\cal M}}
\newcommand{\cN}{{\cal N}}
\newcommand{\cO}{{\cal O}}
\newcommand{\cD}{{\cal D}}
\newcommand{\cL}{{\cal L}}
\def\cR{{\cal R}}
\def\cW{{\cal W}}
\newcommand{\sha}{\, \raisebox{1.2ex}[0mm][0mm]{\rotatebox{270}{$\exists$}} \,}

\newcommand{\vpp}{\mbox{$\langle{\scriptstyle++}\rangle$}}
\newcommand{\vmp}{\mbox{$\langle{\scriptstyle-+}\rangle$}}
\newcommand{\vppp}{\mbox{$\langle{\scriptstyle+++}\rangle$}}
\newcommand{\vmpp}{\mbox{$\langle{\scriptstyle-++}\rangle$}}
\newcommand{\vpmp}{\mbox{$\langle{\scriptstyle+-+}\rangle$}}

\begin{document}
\setlength{\captionmargin}{36pt}
\begin{titlepage}
\begin{flushright}
LAPTH-061/13  \\
\end{flushright}

\vskip 3cm
\begin{center}
\begin{Large}
{\bf Hexagon Wilson Loop OPE\\ and Harmonic Polylogarithms}
\end{Large}

\vskip 2cm
{\large
Georgios Papathanasiou\footnote{E-mail  address: {\tt georgios@lapth.cnrs.fr}}
}
\vskip0.30cm
{\it LAPTh, CNRS, Universit\'e de Savoie, Annecy-le-Vieux F-74941, France}
\vskip0.20cm
and
\vskip0.20cm
{\it  Physics Department, Theory Division, CERN, CH-1211 Geneva 23, Switzerland}

%(\today)

\vskip 1.0cm
\end{center}

\begin{abstract}
\noindent
A recent, integrability-based conjecture in the framework of the Wilson loop OPE for $\mathcal{N}=4$ SYM theory, predicts the leading OPE contribution for the hexagon MHV remainder function and NMHV ratio function to all loops, in integral form. We prove that these integrals evaluate to a particular basis of harmonic polylogarithms, at any order in the weak coupling expansion. The proof constitutes an algorithm for the direct computation of the integrals, which we employ in order to obtain the full (N)MHV OPE contribution in question up to 6 loops, and certain parts of it up to 12 loops. We attach computer-readable files with our results, as well as an algorithm implementation which may be readily used to generate higher-loop corrections. The feasibility of obtaining the explicit kinematical dependence of the first term in the OPE in principle at arbitrary loop order, offers promise for the suitability of this approach as a non-perturbative description of Wilson loops/scattering amplitudes.
\end{abstract}
\vfill
\end{titlepage}

\tableofcontents

\section{Introduction}
The discovery of the duality between Maximally Helicity Violating (MHV) amplitudes and null polygonal Wilson loops in planar $\mathcal{N}=4$ super-Yang Mills theory \cite{Alday:2007hr,Drummond:2007aua,Brandhuber:2007yx} has elucidated remarkable features of its structure, such as dual conformal invariance \cite{Drummond:2006rz} (see also \cite{Drummond:2010km} for a review). The controlled manner in which the latter symmetry is broken, implies that the all-loop behavior of the amplitudes is accurately captured by the BDS ansatz \cite{Bern:2005iz} for four and five points, and needs to be corrected by a scalar function of conformally invariant cross-ratios $u_i$, known as the remainder function $R_n$, for $n=6$ points and beyond \cite{Drummond:2007au}.

For the simplest nontrivial case of six points at two loops $R_6^{(2)}(u_1,u_2,u_3)$, a long expression involving transcendental functions of many variables, known as multiple (or Goncharov) polylogarithms, was first found on the Wilson loop side in \cite{DelDuca:2010zg}. This was then drastically simplified and reexpressed in terms of classical polylogarithms with the method of symbols in \cite{Goncharov:2010jf}. In extracting higher-loop information, it proves advantageous to consider kinematical limits where simplifications occur, such as the multi-Regge \cite{Lipatov:2010ad,Fadin:2011we} and (near-)collinear \cite{Alday:2010ku} limit.

The Operator Product Expansion (OPE) approach to null polygonal Wilson loops is precisely an expansion in terms approaching the collinear limit at different paces, each of which receives contributions at any loop order. For the hexagon, which will be the focus of this paper, it predicts that the leading term at weak 't Hooft coupling $\lambda=g^2_{YM}N$ has the form
\be\label{R_leading_OPE}
R_6=\cos\phi \,e^{-\tau}\sum_{l=1}^\infty \lambda^l \sum_{n=0}^{l-1}\tau^n f^{(l)}_n(\sigma)+\mathcal{O}(e^{-2\tau})\,,
\ee
where $\{\tau,\sigma,\phi\}$ is a particular parametrization of $\{u_1,u_2,u_3\}$, in which $\tau\to\infty$ conveniently describes the limit where two consecutive segments become collinear. As we review in the next section, each term in the expansion corresponds to a different excitation of a color-electric flux tube, created by the two segments adjacent to the ones becoming collinear, whose energy can be calculated exactly \cite{Basso:2010in} with the help of $AdS/CFT$ integrability (see \cite{Beisert:2010jr} for a review).

The functions $f^{(l)}_n(\sigma)$ above are given in terms of a single Fourier integral, whose precise integrand was found in \cite{Alday:2010ku} only for $n=l-1$. Still, this information combined with other reasonable assumptions was enough to fix the 3-loop symbol of the hexagon remainder function up to two unknown parameters \cite{Dixon:2011pw}, which were later determined in \cite{CaronHuot:2011kk}. Apart from the propagation of the flux tube excitation, what was further necessary for obtaining the integrals for any $n$, was knowledge of the transition amplitude, or form factor, describing how the excitation is emitted/absorbed at the two sides of the flux tube.

Great progress in this respect was recently made in \cite{Basso:2013vsa,Basso:2013aha}, where an all-loop expression for the aforementioned form factor, or `pentagon transition', was proposed. This formulation, which now holds for any $n$-gon, again crucially relies on integrability. In particular, it relates the pentagon transition to the S-matrix of excitations on top of the Gubser-Klebanov-Polyakov string \cite{Basso:2013pxa,Fioravanti:2013eia}, which is the string dual to the flux tube vacuum. The integrals of the so called `flattened' part $f^{(l)}_0(\sigma)$ were computed up to $l=4$ loops by means of an ansatz, and perhaps more importantly, it is also possible to obtain more terms in the OPE expansion in this framework \cite{Basso:2014koa}. The data of the leading and subleading OPE contributions were sufficiently strong constraints for determining the full 3-loop hexagon remainder function \cite{Dixon:2013eka}, with the authors of the latter paper stating that this procedure could be extended to four loops and higher.

A practical question that naturally arises in this context, is how one can efficiently evaluate the integrals defining $f^{(l)}_n(\sigma)$ at higher loops. A more conceptual question, is whether there exists a basis of functions large enough to describe the answer at any loop order, especially in light of the conjecture \cite{ArkaniHamed:2012nw}, that multiple polylogarithms should form such a basis for the hexagon Wilson loop in general kinematics.

In this work, we address the two aforementioned questions simultaneously. By identifying the building blocks of the leading OPE integrand, reducing the integral to a sum of residues, and employing the technology of $Z$-sums \cite{Moch:2001zr}, we prove that at any loop order the $\sigma$-dependence of the contribution (\ref{R_leading_OPE}) to $R_6$ is given by
\be
f^{(l)}_n(\sigma)=\sum_{s,r,m_i} c^{\pm}_{s,m_1,\ldots,m_r} e^{\pm\sigma}\sigma^s H_{m_1,\ldots,m_r}(-e^{-2\sigma})\,,
\ee
where the $c^{\pm}$ are numerical coefficients, and $H_{m_1,\ldots,m_r}(x)$ are a single-variable subset of multiple polylogarithms known as harmonic polylogarithms (HPLs) \cite{Remiddi:1999ew}. Our proof is algorithmic in nature, which allows us to perform the integrations in principle at any loop order. We implement it in order to determine $f^{(l)}_n$ for any $n$ up to $l=6$ loops, and for $n=l-1, l-2$ up to $l=12$, thereby providing new high-loop predictions for the MHV hexagon remainder function.

Finally, we also analyze a particular component of the NMHV ratio function $\mathcal{R}$ \cite{Drummond:2008vq}, for which an OPE framework has also been developed \cite{Sever:2011da,Sever:2012qp,Basso:2013aha}. This framework is based on efforts to extend the Wilson loop/amplitude duality beyond the MHV case \cite{Mason:2010yk,CaronHuot:2010ek,Belitsky:2012nu,Belitsky:2012mp,Belitsky:2012rc}, and also on the establishment of an interesting triangle of relations of the latter two observables to correlation functions of operators in the stress tensor multiplet of $\mathcal{N}=4$ theory, in the limit where their consecutive spacetime separations become lightlike \cite{Alday:2010zy,Eden:2010zz,Eden:2010ce,Eden:2011yp,Eden:2011ku}\footnote{In particular, it was shown in \cite{Belitsky:2011zm} that the dimensional regularization of the initial super-Wilson loop proposal \cite{Mason:2010yk,CaronHuot:2010ek} breaks superconformal symmetry, and hence cannot be in correspondence with scattering superamplitudes. More recent studies of the 1- and 2-loop super-Wilson loop suggest that anomalies appear only in certain components \cite{Belitsky:2012mp}, for which the symmetry can be restored either by finite counterterms \cite{Belitsky:2012nu}, or by explicitly rewritting them in an invariant form with the help of the superpropagator \cite{Belitsky:2012rc}. In any case the approach \cite{Sever:2011da,Sever:2012qp,Basso:2013aha} may be thought of as a collinear-limit expansion of the null correlators \cite{Alday:2010zy,Eden:2010zz,Eden:2010ce,Eden:2011yp,Eden:2011ku}.}. We find that at weak coupling
\be
\mathcal{R}^{(6134)}_6= \frac{e^{-\tau}}{2\cosh\sigma}\sum_{l=0}^\infty \lambda^l \sum_{n=0}^{l}\tau^n f^{(l)}_n(\sigma)+\mathcal{O}(e^{-2\tau})\,,
\ee
where for any $l$,
\be
f^{(l)}_n(\sigma)=\sum_{s,r,m_i} c_{s,m_1,\ldots,m_r}\sigma^s H_{m_1,\ldots,m_r}(-e^{-2\sigma})\,,
\ee
and the $c$'s are numerical coefficients. We similarly employ our algorithm implementation in order to obtain explicit expressions for all $f^{(l)}_n$ up to $l=6$, and the $n=l,l-1$ terms up to $l=12$.

This paper is organized as follows. We begin by reviewing the basic ingredients of the OPE approach for the hexagon Wilson loop in section \ref{section_WLReview}, and present the integral formulas describing the leading term for the MHV remainder function and for the component of the NMHV ratio function. In section \ref{section_general_analysis} we prove that their weak coupling expansion at arbitrary loop order always evaluates to the basis of functions described above. Section \ref{section_implementation} focuses on the utility of our proof as a direct evaluation method of these integrals. We first summarize the steps of the algorithm, and then apply it in order to obtain new predictions for the remainder and ratio functions at high loop order. We conclude with an extensive discussion on the implications of our work, and possible directions of further inquiry.

The appendix contains additional information on several functions which were important in our treatment, and most notably harmonic polylogarithms. The results from all our high-loop calculations, as well as the Mathematica code used to generate them, are included in seven ancillary files accompanying the version of this paper on the \texttt{arXiv}.

\section{The Wilson Loop OPE}\label{section_WLReview}
This section serves as a review of the OPE approach for the hexagon Wilson loop, and helps in establishing our notations. In subsection \ref{subsec_OPE_kinematics} we discuss how to take the collinear limit, and outline how at weak coupling the Wilson loop decomposes into terms approaching the limit at different paces, mostly based on \cite{Alday:2010ku,Basso:2010in}. Subsections \ref{subsec_MHV_hexreview} and \ref{subsec_NMHV_hexreview} focus on the extension and refinement of this approach for the MHV and NMHV hexagon respectively, as presented in \cite{Basso:2013vsa,Basso:2013aha}, and also building on \cite{Drummond:2008vq,Sever:2011da,Sever:2012qp}.

For the knowledgeable reader, the equations which will form the basis of our subsequent analysis are the definition of the conformally invariant, finite Wilson loop observable (\ref{rlittle_definition}), and its leading OPE contribution for the MHV case (\ref{r_observable}) and NMHV case (\ref{NMHV_component}), in terms of an all-loop integral.

\subsection{Kinematics and dynamics in the collinear limit}\label{subsec_OPE_kinematics}
\begin{figure}
\centering
\begin{subfigure}[b]{0.49\textwidth}
\includegraphics[width=\textwidth]{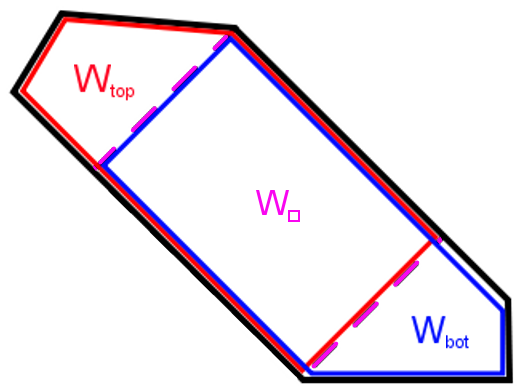}
\caption{\label{fig:hexagon_pieces}}
\end{subfigure}
\begin{subfigure}[b]{0.49 \textwidth}
\includegraphics[width=\textwidth]{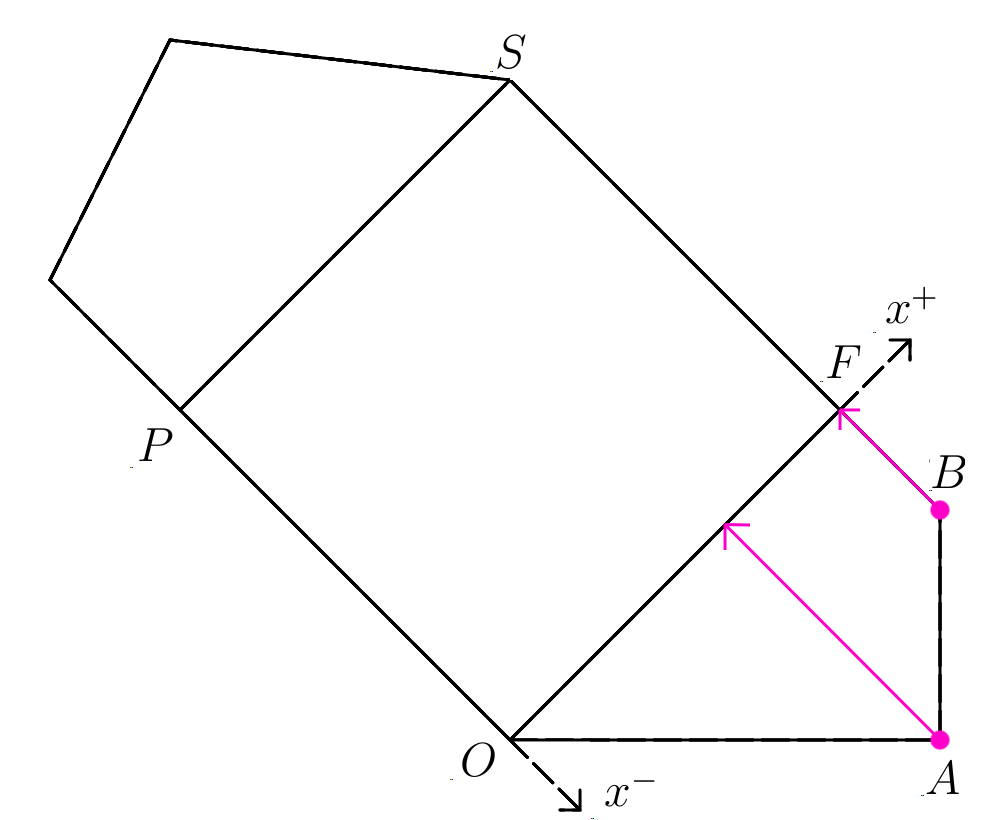}
\caption{\label{fig:hexagon_collinear}}
\end{subfigure}
\caption{\small In (a), the dashed null segments connect two non-intersecting edges of the hexagon, crossing from two of its cusps. They break it up into three squares, with every two adjacent ones forming a pentagon. In (b), using conformal transformations we place $O$ at the origin, and $P, F, S$ at null past, null future, and spacelike infinity respectively. The conformal group element $e^{-\tau(D-M_{01})}$ leaves the middle square invariant, and its action on $A$ and $B$ makes them parallel to $x^+$ when $\tau\to\infty$.}
\label{fig:hexagon}
\end{figure}

In order to take the collinear limit of the hexagon, we start by picking two non-intersecting segments, and form a square by connecting them with another two lightlike segments (see figure \ref{fig:hexagon_pieces}). We can then fix all of its 16 coordinates, 4 of them from the lightlike constraints, and the rest by conformal transformations. Specifically, the 4-dimensional conformal group $SO(2,4)$ will have 15 generators, which implies that not only all squares will be conformally equivalent, but also that each given square will be invariant under a subset of 3 transformations.

As we show in figure \ref{fig:hexagon_collinear}, a convenient choice will be to arrange the square to lie on the $(x^0,x^1)$ plane, with its points located at the origin, past lightlike infinity ($x^-=x^0-x^1\to-\infty$ with $x^+=x^0+x^1$ fixed), future lightlike infinity ($x^+\to\infty$ with $x^-$ fixed), and spacelike infinity ($x^1\to\infty$ with $x^0$ fixed). In this case, the three symmetries of the square will be dilatations, boosts in the $(x^0,x^1)$ plane, and rotations in the transverse $(x^2,x^3)$ plane, generated by $D$, $M_{01}$ and $M_{23}$ respectively.

For concreteness, we may choose the generators and one-parameter group elements to be
\begin{align}
M_{01}&=\begin{pmatrix}
0&1&0&0\\
1&0&0&0\\
0&0&0&0\\
0&0&0&0
\end{pmatrix}\,,& e^{-\xi M_{01}}&=\begin{pmatrix}
\cosh\xi&-\sinh\xi&0&0\\
-\sinh\xi&\cosh\xi&0&0\\
0&0&1&0\\
0&0&0&1
\end{pmatrix}\,,\\
M_{23}&=\begin{pmatrix}
0&0&0&0\\
0&0&0&0\\
0&0&0&-i\\
0&0&i&0
\end{pmatrix}\,,& e^{-i \phi M_{23}}&=\begin{pmatrix}
1&0&0&0\\
0&1&0&0\\
0&0&\cos\phi&-\sin\phi\\
0&0&-\sin\phi&\cos\phi
\end{pmatrix}\,,
\end{align}
whereas $D$ is just the identity matrix, $e^{\lambda D}=e^\lambda$. It is straightforward to show their action on the square leaves it invariant, also keeping in mind that in two dimensions spacelike infinity is a single point.

Going back to the hexagon, we can parametrize all conformally inequivalent geometries by acting with the symmetries of the square on the cusps below the reference square. In fact, it will be advantageous to consider the group element $\exp\left[-\tau(D-M_{01})\right]$, as the collinear limit will now simply correspond to $\tau\to\infty$. In more detail,
\be
\left(\lim_{\tau\to\infty}e^{-\tau(D-M_{01})}\right)\cdot x^\mu=\begin{pmatrix}
\frac{1}{2}&\frac{1}{2}&0&0\\
\frac{1}{2}&\frac{1}{2}&0&0\\
0&0&0&0\\
0&0&0&0
\end{pmatrix}\cdot
\begin{pmatrix}
x^1\\
x^2\\
x^3\\
x^4
\end{pmatrix}=\frac{1}{2}(x^0+x^1)\begin{pmatrix}
1\\
1\\
0\\
0
\end{pmatrix}\,.
\ee
This implies that all segments below the middle square will flatten out onto its lower edge, as we illustrate in figure \ref{fig:hexagon_collinear}, thus becoming collinear. Depending on the choice of initial hexagon, we will get a slightly different relation between the cross ratios $u_i$ and the group coordinates $\tau, \sigma, \phi$ parametrizing the symmetries of the square. Following the conventions of \cite{Basso:2013vsa,Basso:2013aha}, we will be using\footnote{Note that this parametrization differs from the one used in the initial Wilson loop OPE approach \cite{Alday:2010ku}.}
\be
\begin{aligned}
u_1&=\frac{x_{46}^2x_{13}^2}{x_{36}^2 x_{14}^2} =\frac{1}{2}\frac{e^{2 \sigma+\tau}  \text{sech}\tau}{1+e^{2 \sigma }+ 2\,e^{\sigma -\tau } \cos \phi +e^{-2 \tau }}\,,\\
u_2&=\frac{x_{15}^2 x_{24}^2}{x_{14}^2 x_{25}^2}= \frac{1}{2} e^{-\tau } \text{sech}\tau  \,,  \\
u_3&=\frac{x_{26}^2 x_{35}^2}{x_{25}^2x_{36}^2} = \frac{1}{1+e^{2 \sigma }+ 2\,e^{\sigma -\tau } \cos \phi +e^{-2 \tau }}  \,. \label{crossratios}
\end{aligned}
\ee

A great advantage of describing the collinear limit in a matter which takes into account the symmetries of the square, is that it also makes the description of the dynamics more transparent. In particular, we can think of the Wilson loop segments belonging to the middle square as a flux tube sourced by two quarks moving at the speed of light, and decompose the Wilson loop with respect to all possible excitations of this flux tube. These excitations will be eigenstates of the symmetries of the square with eigenvalues $E, p, \phi$, so that schematically we may write
\be\label{OPE_schematic}
W=\int dn e^{-\tau E_n+i p_n+i m_n\phi}\, C_{\text{bot}}C_{\text{top}}\,,
\ee
where the exponential part describes their propagation, $n$ labels different excitations, and $C_{\text{bot}}$ ($C_{\text{top}}$) denotes the transition amplitude, or overlap, between the initial (final) state of the bottom (top) part of the polygon and the intermediate eigenstate. This picture is reminiscent of expressing the product of two neighboring operators $A$ and $B$ as a sum of local operators, whose scaling dimensions control the dependence of the coefficients on the distance between $A$ and $B$. Hence the decomposition (\ref{OPE_schematic}) around the collinear limit has been coined `Wilson loop OPE'.

Since its derivation only relied on the symmetries of the problem, the above formula should hold for any conformal field theory where the flux is conserved. The good news is that in $\mathcal{N}=4$ super Yang-Mills, the flux tube excitations are in 1-1 correspondence with excitations of an integrable spin chain, with the collinear twist operator \cite{Braun:2003rp} $D-M_{01}$ as its hamiltonian. The states of the spin chain are single-trace operators, with the vacuum made of a sea of derivatives acting on the complex combination of two scalars of the theory $Z$,
\be
\text{vacuum}=\text{tr}\left(ZD_+^SZ\right)\,,\quad D_+=D_0+D_1
\ee
and excited states built by inserting any fundamental field of the theory $\Phi$ on the vacuum, e.g.
\be
\text{single excitation}=\text{tr}\left(ZD_+^{S_1}\Phi D_+^{S_2}Z\right)\,,\label{GKP_excitation}
\ee
where $S\sim S_1+S_2\gg0$. In fact, minimizing the energy suggests that elementary excitations of the spin chain can only be drawn from the components of $\Phi$ which have minimal classical twist eigenvalue $\Delta-S=1$. At quantum level the (shifted with respect to the vacuum) twist of the operators (\ref{GKP_excitation}) receives anomalous contributions due to renormalization,
\be
E(p)=(\Delta-S)_1-(\Delta-S)_{\text{vac}}=1+\sum_{l=1}^\infty \lambda^l E^{(l)}(p)\,.
\ee

Due to integrability, the `energy' $E$ can be calculated to all loops \cite{Basso:2010in}, and moreover for $M$ elementary excitations we will have $E_M=M+\mathcal{O}(\lambda)$. This teaches us that at weak coupling the Wilson loop OPE (\ref{OPE_schematic}) will be a sum of terms with different integer exponential behaviors $e^{-\tau M}$ as $\tau\to\infty$, where $M$ is the classical twist of the state, or equivalently the number of elementary excitations it consists of. In this paper, we will focus on the leading-twist, or single-particle term $\mathcal{O}(e^{-\tau})$, and we will frequently refer to it as the `leading OPE contribution'.

\subsection{MHV hexagon}\label{subsec_MHV_hexreview}

Let us now become more specific, and define the hexagon Wilson loop-related observable which will be most suited for analyzing its OPE. By construction, the middle square always has two of its segments coinciding with the cusps of the hexagon, and hence the bottom and top part of the hexagon which lie outside of it will also be squares. Then,
\be\label{rlittle_definition}
r\equiv\log \mathcal{W}\equiv\log\frac{W W_{\Box}}{W_{\text{bot}}W_{\text{top}}}\,,
\ee
where $W$ is our hexagon, and $W_{\Box}, W_{\text{bot}}, W_{\text{top}}$ are the Wilson loops defined on the contours of the middle square, and of the pentagons created by joining the middle and lower, and middle and upper squares (see figure \ref{fig:hexagon_pieces}). The particular ratio we are considering removes all (cusp-induced) ultraviolet divergences, leaving a finite function of conformal cross ratios. Aside this, it does not cause any loss of information, as the square and pentagon Wilson loops are given by the BDS ansatz \cite{Bern:2005iz}.

Specializing on the MHV case, from symmetry arguments the single-particle contribution is expected to be bosonic and uncharged under the R-symmetry. This picks out only one out of the twist-1 excitations, the component $F_{+i}$ of the gauge field \cite{Alday:2010ku}, where the first component is projected on $x^+$ and the second component is on the $(x^3,x^4)$ plane, according to the notations discussed in the beginning of section \ref{subsec_OPE_kinematics}. As we also mentioned in that section, its all-loop dispersion relation has been found with the help of integrability \cite{Basso:2010in}, in parametric form with respect to the Bethe rapidity $u$. Following the notations of the latter paper, from this point on we will rescale the 't Hooft coupling $\lambda$ by
\be
g^2\equiv\frac{\lambda}{(4\pi)^2}\,,
\ee
in terms of which the gauge field dispersion relation reads
\be\label{Ep_gluon}
\begin{aligned}
\gamma_1(u)&\equiv E_1(u)-1 = \int_{0}^{\infty}{dt \over t}\bigg[{\gamma^{\varnothing}_{+}(2gt) \over 1-e^{-t}} -{\gamma^{\varnothing}_{-}(2gt) \over e^{t}-1}\bigg]\left(\cos{(ut)}e^{- t/2}-1\right)\, , \\
p_1(u) &= 2u-\int_{0}^{\infty}{dt \over t}\bigg[{\gamma^{\varnothing}_{-}(2gt) \over 1-e^{-t}} +{\gamma^{\varnothing}_{+}(2gt) \over e^{t}-1}\bigg]\sin{(ut)}e^{- t/2}\, ,
\end{aligned}
\ee
where the functions $\gamma^{\varnothing}_{\pm}(2gt)$ are independent of $u$, and as we review in appendix \ref{appx_gammao_functions}, can be obtained iteratively as a Taylor expansion in $g\ll1$.

In order to obtain information about the hexagon from the analogue of (\ref{OPE_schematic}) for $r$, we need however to know the (rescaled) creation/absorption form factors $C_{\text{bot}}, C_{\text{top}}$, which also depend on $g$. Initially, these were determined to the first few orders by comparing with the explicit computation of the hexagon up to two loops \cite{DelDuca:2010zg,Goncharov:2010jf}. Recently however, all-loop expressions for them were proposed, relying again on the integrability of the theory \cite{Basso:2013vsa,Basso:2013aha}. In particular, these form factors, also dubbed as `pentagon transitions', are related to the S-matrix of excitations on top of the GKP string \cite{Basso:2013pxa,Fioravanti:2013eia}, which forms the vacuum of the flux tube.\footnote{The new formulation in the framework of the OPE approach also has the advantage of being applicable to Wilson loops with more cusps, and generalizable to multiparticle contributions \cite{Basso:2014koa}.}

With this ingredient in place, and in the more natural rapidity parametrization, the final formula for the leading (single-particle) OPE contribution for $r$ reads
\be\label{r_observable}
r=2\cos\phi e^{-\tau} \int_{-\infty}^{+\infty} \frac{du}{2\pi}  \mu_1(u) e^{-\gamma_1(u)\tau+i p_1(u)\sigma} + \mathcal{O}(e^{-2\tau}),
\ee
where the measure $\mu_1(u)$ is given by
\be\label{measure_gluon}
\begin{aligned}
&\mu_1(u) =-\frac{\pi g^2}{\cosh{(\pi u)}}\frac{\left(u^2+\frac{1}{4}\right)}{(x^{+}x^{-}-g^2)\sqrt{(x^+x^{+}-g^2)(x^-x^{-}-g^2)}}\times\\
&\exp{\bigg[\int\limits_{0}^{\infty}\frac{dt}{t}(J_{0}(2gt)-1)\frac{2e^{-t/2}\cos(ut)-J_{0}(2gt)-1}{e^{t}-1} +f_3(u, u)-f_4(u, u)\bigg]}\, .
\end{aligned}
\ee
In the last formula, $x^\pm$ are the Zhukowski variables
\be
x^{\pm} = x(u\pm \ft{i}{2})\,,\quad x(u) = \frac{u+\sqrt{u^2-(2g)^2}}{2}\,,
\ee
$J_{i}$ is the $i$-th Bessel function of the first kind, and the form of the functions $f_i(u, v)$ is reviewed in appendix \ref{appx_f_functions}. Once the observable (\ref{r_observable}) has been determined, the near collinear limit of the MHV hexagon remainder function $R$ (we drop the index as we will only be dealing with $n=6$),
\be\label{R_definition}
W=W^{\text{BDS}}e^{R(u_1,u_2,u_3)}
\ee
parametrizing the part of the Wilson loop $W$ which is not captured by the BDS ansatz $W^{\text{BDS}}$ \cite{Bern:2005iz}, is given by
\be\label{r_to_R}
R=r-r^{BDS}\,,
\ee
where \cite{Gaiotto:2011dt}
\begin{align}
r^{BDS}\equiv\log\cW^\text{BDS}&={\Gamma_\text{cusp}\over4}\{\text{Li}_2\left(u_2\right)-\text{Li}_2\left(1-u_1\right)-\text{Li}_2\left(1-u_3\right)+\log ^2\left(1-u_2\right)\nonumber\\
&\quad-\log \left(u_1\right) \log\left(u_3\right)-\log\left(u_1/ u_3\right)\log\left(1-u_2\right)+\frac{\pi ^2}{6}\}\\
&=-{\Gamma_\text{cusp}\over2}\cos\phi  e^{-\tau}\left[e^{-\sigma } \log\left({1+e^{2 \sigma }}\right)+ e^{\sigma } \log\left(1+e^{-2 \sigma }\right)\right]+\mathcal{O}(e^{-2\tau})\,,\label{r_BDS}
\end{align}
is defined as in (\ref{rlittle_definition}), but only including the BDS contribution for each of the polygons. The equality (\ref{r_to_R}) is a consequence of the definition (\ref{R_definition}), given that the BDS ansatz accurately describes square and pentagon Wilson loops.

In more detail, the logarithm of the BDS part of any null polygonal Wilson loop in $\mathcal{N}=4$ super Yang-Mills, is proportional to the logarithm of the same Wilson loop in the $U(1)$ theory, with the proportionality constant being a fourth of the cusp anomalous dimension $\Gamma_\text{cusp}$. The latter quantity can also be calculated to all loops (see \cite{Freyhult:2010kc} for a review), and is given to the first few orders in the weak coupling by
\be\label{gamma_cusp}
\Gamma_\text{cusp}=\sum_{l=1}^\infty g^{2l}\Gamma_\text{cusp}^l=4 g^2-\frac{4\pi ^2}{3} g^4 +\frac{44 \pi ^4}{45}g^6-4 \left(\frac{73 \pi ^6}{315}+8 \zeta_3^2\right)g^8+\mathcal{O}(g^{10})\,.
\ee

\subsection{NMHV hexagon}\label{subsec_NMHV_hexreview}

Scattering amplitudes involving gluons in other helicity configurations, or other particles of $\mathcal{N}=4$ super Yang-Mills, are most conveniently described by exploiting its (dual) superconformal symmetry \cite{Drummond:2008vq}, see also \cite{Drummond:2010km} for a review.

Very briefly, one starts by packaging the particle content of the theory\footnote{The on-shell fields of $\mathcal{N}=4$ super Yang-Mills form the CPT self-conjugate, `doubleton' representation of the superconformal group $PSU(2,2|4)$ \cite{Gunaydin:1984fk}.} in a single superfield $\Phi$ with the help of a Grassmann variable $\eta^A$, whose index transforms in the fundamental representation of the R-symmetry group $SU(4)$. Namely, all external states of $\pm1$ helicity gluons $G^\pm$, $\pm\frac{1}{2}$ helicity Majorana fermions $\Gamma_A, \bar \Gamma^A$, and zero helicity real scalars $S_{AB}$ can be simultaneously described by
\be
\Phi=G^+ +\eta^A\Gamma_A+\tfrac{1}{2!}\eta^A\eta^B S_{AB}+\tfrac{1}{3!}\eta^A\eta^B\eta^C\epsilon_{ABCD}\bar \Gamma^D+\tfrac{1}{3!}\eta^A\eta^B\eta^C\eta^D\epsilon_{ABCD}G^-\,,
\ee
which in turn allows us to combine all $n$-point amplitudes in a superamplitude $\mathcal{A}_n(\Phi_1,\ldots,\Phi_n)$.

All MHV amplitudes form the part of the superamplitude which has 8 powers of Grassmann variables, starting the the MHV gluon amplitude,
\begin{align}\label{gen-MHV}
{\cal A}_{n}^{\rm MHV} =(2\pi)^4 \delta^{(4)} \big(\sum_{i=1}^n p_i\big)\sum_{1\le j<k\le n}
(\eta_j)^4  (\eta_k)^4 A^{\rm MHV} _n(1^+...\, j^-...\,  k^-...\,  n^+) + \ldots\,.
\end{align}
On the basis of tree- and 1-loop level amplitude computations, it was argued in \cite{Drummond:2008vq} that NMHV amplitudes will similarly organize in a homogeneous polynomial of degree $12$ in $\eta^A_i$, and more importantly, that they have the same infrared divergence structure as the MHV amplitudes, so that the two superamplitudes are related by
\be
\mathcal{A}^{\text{NMHV}}_n=\mathcal{A}^{\text{MHV}}_n\mathcal{R}_n\,,
\ee
where $\mathcal{R}_n$ is the (dual conformal invariant) NMHV ratio function. Evidently, it will consist of terms involving 4 powers of the Grassmann variables, whose components we can be denoted as $\mathcal{R}^{(ijkl)}_n$.

Following attempts for generalizing the Wilson loop/scattering amplitude duality beyond the MHV case \cite{Mason:2010yk,CaronHuot:2010ek}, an analogous proposal for the OPE of certain components of the NMHV hexagon Wilson loop was put forth in \cite{Sever:2011da,Sever:2012qp}. According to the latter, the dual of the $\mathcal{R}^{(i,i+1,j,j+1)}_6$ component is given by a Wilson loop\footnote{More precisely, $\mathcal{R}_6$ and $\mathcal{W}_6$ will only differ by $\mathcal{O}(e^{-2\tau})$ terms \cite{Basso:2013aha}.} $\mathcal{W}^{(i,i+1,j,j+1)}_6$ - normalized by bosonic squares and pentagons as in (\ref{r_observable}) - which at tree-level has insertions of a complex scalar field combination at the cusp between segments $i,i+1$, and its complex conjugate field at the cusp between segments $j,j+1$. The two segments forming the flux tube reference square are chosen to lie between $i,i+1$ and $j,j+1$ without coinciding with any of them, and $|i-j|\ge3$. Thus the top and the bottom part of the polygon have a scalar insertion each, and it is also expected that only scalar excitations of the flux tube will propagate to leading order in the OPE expansion.

For a particular component of the NMHV ratio function, this leading OPE contribution was also predicted in \cite{Basso:2013aha}, by a similar analysis of the scalar pentagon transitions. It is the component with the scalar insertions between edges 6-1 and 3-4, and edges 2 and 5 forming the reference frame, for which the OPE reads (we drop the lower index)
\be\label{NMHV_component}
\mathcal{W}^\text{(6134)}=\frac{1}{g^2}\,e^{-\tau} \int \frac{du}{2\pi}  \mu_0(u)  \,e^{-\tau \gamma_0 (u)+i p_0(u)\sigma} +\mathcal{O}(e^{-2\tau})\,,
\ee
where $\gamma_0(u)$ and $p_0(u)$ are the anomalous part of the energy and the momentum of a scalar excitation propagating between the bottom and top part of the polygon \cite{Basso:2010in},
\be\label{Ep_scalar}
\begin{aligned}
\gamma_0(u) &= \int_{0}^{\infty}{dt \over t}{\gamma^{\varnothing}_{+}(2gt)-\gamma^{\varnothing}_{-}(2gt) \over e^{t}-1}\left(e^{t/2}\cos{(ut)}-1\right)-\int_{0}^{\infty}{dt \over t}\gamma^{\varnothing}_{+}(2gt)\, , \\
p_0(u) &= 2u-\int_{0}^{\infty}{dt \over t}{\gamma^{\varnothing}_{+}(2gt)+\gamma^{\varnothing}_{-}(2gt) \over e^{t}-1}e^{t/2}\sin{(ut)}\, ,
\end{aligned}
\ee
and the scalar integration measure $\mu_0$ is now given by
\be\label{measure_scalar}
\begin{aligned}
\mu_0(u)& =\frac{\pi g^2}{\cosh{(\pi u)}}\times\\
&\quad\exp{\bigg[\int\limits_{0}^{\infty}\frac{dt}{t}(J_{0}(2gt)-1)\frac{2e^{t/2}\cos(ut)-J_{0}(2gt)-1}{e^{t}-1} + f_3(u, u)-f_4(u, u)\bigg]}\, .
\end{aligned}
\ee
Here as well $J_{i}$ is the $i$-th Bessel function of the first kind, $\gamma^{\varnothing}_{\pm}$ are the same functions which appeared in the MHV hexagon subsection, and which are reviewed in appendix \ref{appx_gammao_functions}, and $f_i(u, v)$ are functions which are in turn reviewed in appendix \ref{appx_f_functions}. We'll also have the same parametrization (\ref{crossratios}) of the conformally invariant cross ratios $u_1, u_2, u_3$ in terms of $\tau, \sigma, \phi$.

\section{General Analysis of the Integrals}\label{section_general_analysis}
In section \ref{section_WLReview}, we recalled that the Wilson loop OPE approach yields the leading, $\mathcal{O}(e^{-\tau})$ term in the $\tau\to\infty$ collinear limit of the (N)MHV hexagon as a 1-dimensional Fourier integral, to all orders in $g$. Here we will focus on the weak coupling expansion of the integral, and prove that it can be evaluated at any loop order, in terms of harmonic polylogarithms.

\subsection{Reduction to sum over residues}
Let us start by combining the descriptions of the MHV and NMHV integrals (\ref{r_observable}), (\ref{NMHV_component}) simultaneously\footnote{We drop the indices in $\mu_\eta, \gamma_\eta, p_\eta$ in order to avoid clutter, but we should keep in mind they correspond to different functions for $\eta=0,1$.},
\be\label{main_integral}
I_\eta=\int \frac{du}{2\pi}  \mu(u)  \,e^{-\tau \gamma (u)+i p(u)\sigma}\,,
\ee
with the help of the definition
\be\label{mu_combined}
\begin{aligned}
&\mu(u) =\frac{\pi g^2}{\cosh{(\pi u)}}\left(-\frac{x^{+}x^{-}}{(x^{+}x^{-}-g^2)\sqrt{(x^+x^{+}-g^2)(x^-x^{-}-g^2)}}\right)^\eta\times\\
&\exp{\bigg[\int\limits_{0}^{\infty}\frac{dt}{t}(J_{0}(2gt)-1)\frac{2e^{t/2}\cos(ut)-J_{0}(2gt)-1}{e^{t}-1} +f_3(u, u)-f_4(u, u)\bigg]}\,,
\end{aligned}
\ee
where $\eta=1$ corresponds to the propagation of a gauge field excitation and describes the MHV case, and $\eta=0$ corresponds to a scalar excitation and describes the NMHV case. Similarly $\gamma, p$ and the functions $f_3, f_4$ will implicitly depend on $\eta$, as reviewed in appendix \ref{appx_f_functions}, equations (\ref{f_definitions}), (\ref{kappa_vectors}) and (\ref{ep_from_QM}). In obtaining (\ref{mu_combined}), we made use of the identity
\be
\int_0^\infty\frac{dt}{t}2e^{-t/2} \left[1-J_0(2 g t)\right] \cos(t u)=\log\frac{x^+ x^-}{u^2+\frac{1}{4}}\,,
\ee
which may be proven as was done in \cite{Eden:2006rx,Basso:2010in} for similar integrals involving Bessel functions\footnote{We thank Benjamin Basso for pointing this out to us.}, in order to replace $e^{-t}\to e^{t}$ in the second line of (\ref{measure_gluon}).

We will proceed to evaluate the integral (\ref{main_integral}) order by order at weak coupling $g$ by turning it into a sum over residues. To this aim, we will need to analyze the general dependence of $\mu, \gamma$ and $p$ on the integration parameter $u$.

Focusing first on the exponential part of $\mu$, and in particular the $f_i$ functions contained therein, and defined in (\ref{f_definitions}), it is evident that the dependence on $u$ enters only through the $\kappa, \tilde \kappa$ vectors (\ref{kappa_vectors}). As a result of the regular Taylor expansion of the Bessel functions involved (\ref{BesselJ_expansion}), at weak coupling these can always be expressed in terms of the following integrals which can be calculated exactly\footnote{In particular the $m=0$ case only appears in $\kappa_1$, and as the difference $\psi(z)-\psi(1)$, so that the second term in the integral cancels out.},
\be\label{psi_integral}
\int_0 ^\infty dt\left(\frac{t^m e^{-zt}}{1-e^{-t}}-\frac{e^{-t}}{t}\delta_{m,0}\right)=(-1)^{m+1}\psi^{(m)}(z)\,,\quad m\ge0\,,
\ee
where $\psi^{(m)}(z)$ is the polygamma function. It is easy to see that in both NMHV ($\eta=0$) and MHV ($\eta=1$) case the possible arguments include $z=\tfrac{1}{2}\pm i u$ and $z=1$, and for the second argument we further reduce to Riemann zeta functions,
\be
\psi^{(m)}(1)=(-1)^{m+1}m!\zeta_{m+1}\,.
\ee
For the MHV case, there exists one additional possibility for the argument, $z=\tfrac{3}{2}\pm i u$, and by definition the weak coupling expansion of the $f_i$ functions will contain bilinear combinations of these $\psi$-functions.

A similar analysis can be performed for the remaining exponential part of $\mu(u)$ (\ref{mu_combined}), and also for $\gamma(u)$ and $p(u)$, which shows that we now obtain monomials of the aforementioned polygamma functions with the specific arguments. For $\gamma(u), p(u)$ this is particularly easy to see due to the alternative expression (\ref{ep_from_QM}) mentioned in appendix \ref{appx_f_functions}, which relates them to the same building blocks of the $f_i$ functions.

Finally, it is straightforward to show that at any loop order in $g\ll1$, the $x^{\pm}$-dependent factor of the measure (\ref{mu_combined}) will be a sum of products of inverse powers of $u\pm \tfrac{i}{2}$. In more detail, the factor in question is equal to $(u^2+\tfrac{1}{4})^{-1}$ at $g=0$, has a regular Taylor expansion around that point, and due to\footnote{We have chosen the branch $\sqrt{u^2}=u$, which is equivalent to assuming $x(u)$ starts at $\mathcal{O}(g^0)$. The opposite branch corresponds to $x(u)$ starting at $\mathcal{O}(g^2)$, and the two solutions are related by  $x(u)\to g^2/x(u)$, as can be seen by the defining relation $u=x(u)+g^2/x(u)$. From this it follows that the expansions in the two branches of the factor in question only differ by an overall sign, but as pointed out in \cite{Basso:2013aha}, there already exists a sign ambiguity in $\mu(u)$, that needs to be fixed by physical input.}
\be
x^{\pm} = x(u\pm \ft{i}{2})\,,\quad x(u) = \frac{u+\sqrt{u^2-(2g)^2}}{2}=\frac{u}{2}\left[1+\sum_{k=0}^\infty \binom{1/2}{k}\left(\frac{-4g^2}{u^2}\right)^k\right]\,,
\ee
it has effective expansion parameters $g/(u\pm \tfrac{i}{2})$.

Gathering the information we obtained, we deduce that the weak coupling expansion of the integral (\ref{main_integral}) will be a sum of the general form
\begin{align}
&I_\eta=\int \frac{du}{2\pi}  \mu(u)  \,e^{-\tau \gamma (u)+i p(u)\sigma}=\sum_{l=1}^\infty g^{2l}\sum_{n=0}^{l-1}\tau^n \int du \tilde h^{(l)}_n(u,\sigma)\equiv\sum_{l=1}^\infty g^{2l}\sum_{n=0}^{l-1}\tau^n h^{(l)}_n(\sigma)\,,\label{hex_integral}\\
&\tilde h^{(l)}_n(u,\sigma)= \sum c \,e^{2i u\sigma}\text{sech}(\pi u) \prod_{i} \psi^{(m_i)}(\frac{1}{2}\pm i u)\left(\frac{\prod_{j}\psi^{(m_j)}(\frac{3}{2}\pm i u)}{(u+\frac{i}{2})^{r_1} (u-\frac{i}{2})^{r_2}}\right)^\eta\,\label{htilde_genform1}
\end{align}
where we have absorbed all factors that don't depend on $u$ in the coefficients $c$ (which obviously also depend on the different indices $l_i, m_i, r_i$ etc).

From the last formula, we arrive at the following important conclusion: \emph{The only possible locations where the integrand may have poles are for} $u=(k+\frac{1}{2})i$, $k\in \mathbb{Z}$. In particular these poles may come from the denominator (if any), the hyperbolic cosecant, or the polygamma functions at negative integer arguments.

In what follows we will restrict to $\sigma>0$, in which case we can close the contour with a semicircle on the $u>0$ plane, whose integral at infinite radius will go to zero due to the decaying exponential. Therefore by Cauchy's residue theorem we will have
\begin{equation}\label{int_to_residues}
h^{(l)}_n(\sigma)\equiv\int du \tilde h^{(l)}_n(u,\sigma)=2\pi i \sum_{k=0}^\infty \text{Res}\left(\tilde h^{(l)}_n,u=(k+\tfrac{1}{2}) i\right)\,,
\end{equation}
and in order to proceed we will need to find an analytic expression for the residues as a function of $k$. This cannot be achieved by directly Taylor expanding the expressions for the integrands around $u=(k+\frac{1}{2})i$, as $\psi^{(n)}\left(\tfrac{3}{2}+i u\right)$ and $\psi^{(n)}\left(\tfrac{1}{2}+i u\right)$ develop poles there. Instead, we first employ the reflection formula
\be\label{reflection_formula}
\psi ^{(n)}(z)=(-1)^{n} \psi ^{(n)}(1-z)-\pi  \frac{\partial ^n}{\partial z^n}\cot (\pi  z)\,,
\ee
which allows us isolate the singular terms into elementary functions with known expansions. In more detail, we first use the recurrence relation
\be\label{psi_recurrence}
\psi ^{(n)}(z+1)=\psi ^{(n)}(z)+(-1)^n n! z^{-n-1}\,,
\ee
with $z=\frac{1}{2}\pm i u$, in order to eliminate the $\psi^{(m)}(\frac{3}{2}\pm i u)$ factors in (\ref{htilde_genform1})\footnote{Notice that after expanding, we will not get any new rational factors apart from the already existing $u\pm\frac{i}{2}$ in (\ref{htilde_genform1}).}. Then, we apply (\ref{reflection_formula}) with $z=\frac{1}{2}+i u$, also noting that the cotangent derivative will give a polynomial in cotangents and cosecants, which in terms of $u$ become $-i\tanh \pi u$ and $-\text{sech} \pi u$ respectively\footnote{In more detail, it can be shown that
\[
\frac{\partial ^n}{\partial z^n}\cot (\pi  z)=\sum_{k=0}^{\left[\frac{m-1}{2}\right]}c_k \cos(\pi z)^{m-1-2k}\sin(\pi z)^{-m-1}\,,
\] which implies we will always have even powers of $\csc(\pi z)\to -\text{sech} \pi u$.\label{footnote_dcos}}

In this manner, we have achieved to reexpress (\ref{htilde_genform1}) as a sum of products of a restricted set of functions, having known expansions for $u=(k+\tfrac{1}{2})i+\epsilon$, $\epsilon$ small, which we also present below:
\begin{align}
e^{2iu \sigma}&= e^{-(2k+1)\sigma+i\epsilon\sigma}=e^{-(2k+1)\sigma}\sum_{n=0}^{\infty}\frac{(i\epsilon\sigma)^n}{n!}\,,\label{exponential_series}\\
\psi ^{(m)}\left(\tfrac{1}{2}-i u\right)&=\psi ^{(m)}\left(k+1-\tfrac{i\epsilon}{2}\right)=\sum_{n=1}^\infty \frac{\psi^{(m+n)}(k+1)}{n!}\left(\frac{-i\epsilon}{2}\right)^n\,,\label{psi_expansion}\\
\tanh \pi u&=\coth \frac{\epsilon\pi}{2}=\frac{2}{\epsilon\pi}+\sum _{n=1}^{\infty } \frac{2^{2 n} B_{2 n} }{(2 n)!}\left(\frac{\epsilon\pi}{2}\right)^{2 n-1}\,,\label{tanh_expansion}\\
\text{sech}\pi u&=-i (-1)^k \text{csch}\frac{\epsilon\pi  }{2}=-i (-1)^k\left[\frac{2}{\epsilon\pi}-\sum _{n=1}^{\infty } \frac{2 \left(2^{2 n-1}-1\right) B_{2 n}}{(2 n)!}\left(\frac{\epsilon\pi  }{2}\right)^{2 n-1}\right]\,,\label{sech_expansion}\\
\frac{1}{u-\frac{i}{2}}&=\frac{-i}{k+1-\frac{i\epsilon}{2}}=\frac{-i}{k+1}\sum_{n=0}^\infty \left(\frac{i\epsilon}{2(k+1)}\right)^n\,,\label{special_fraction_series_one}\\
\frac{1}{u+\frac{i}{2}}&=\frac{-i}{k-\frac{i\epsilon}{2}}=\begin{cases}
\frac{2}{\epsilon}&k=0\,,\\
-\frac{i}{k}\sum_{n=0}^\infty \left(\frac{i\epsilon}{2k}\right)^n&k\ge1\,,
\end{cases}\label{special_fraction_series}
\end{align}
where $B_{2 n}$ are the Bernoulli numbers.

A few remarks are in order. First, since the hyperbolic functions become periodic in the imaginary axis, their $k$-dependence reduces to an overall sign at most. Furthermore, because of the last equation, we will have to determine the residue for $k=0$ separately. Aside this value, we may use the above expansions in order to determine the residues in (\ref{int_to_residues}) for general positive integer $k$. Finally, the fact that the residues will only contain $\psi$-functions with integer arguments, allows us to replace them with generalized harmonic numbers,
\be\label{psi_to_H}
\begin{aligned}
\psi(k+1)&\equiv\psi^{(0)}(k+1)=-\gamma_E+S_1(k)\\
\psi^{(m-1)}(k+1)&=(-1)^{m} (m-1)! (\zeta_{m}-S_{m}(k))\,,
\end{aligned}
\ee
defined as
\be\label{HarmonicN_definition}
S_{m}(k)=\sum _{n=1}^k \frac{1}{n^m}\,,
\ee
where $\gamma_E=-\psi(1)=\simeq 0.577$ is the Euler-Mascheroni constant.

\subsection{$Z$-sums and Harmonic Polylogarithms}

Let us now focus on the structure of the residue of $\tilde h^{(l)}_n$ at $u=(k+\tfrac{1}{2})i$, as a function of $k$. In the previous section, we demonstrated that $\tilde h^{(l)}_n$ will be a sum of products of the restricted set of functions (\ref{psi_expansion})-(\ref{psi_to_H}), which we expanded around the positions of all poles. From this analysis, it immediately follows that
\be\label{generic_residue_term}
\text{Res}\left(\tilde h^{(l)}_n,u=(k+\tfrac{1}{2}) i\right)= \sum c \,e^{-\sigma} (-e^{-2\sigma})^{k}\sigma^s\left(\frac{1}{k^{l_1} (k+1)^{l_2}}\right)^\eta\prod_{i=1}^r S_{m_i}(k)\,
\ee
for different values of the indices $s,r, l_1,l_2, m_1,\ldots m_r$ and numerical constants $c$.

In more detail, (\ref{tanh_expansion}) will only contribute numerical factors independent of $k$. The same will apply for (\ref{sech_expansion}), except for an overall $(-1)^k$ factor, as the discussion in footnote {\ref{footnote_dcos} implies that we will always have odd powers of $\text{sech}\pi u$. The latter $k$-dependent factor combines with the exponential coming from (\ref{exponential_series}), which will also contribute the powers of $\sigma$. Finally the inverse powers of $k, k+1$ come from (\ref{special_fraction_series_one}), (\ref{special_fraction_series}), and the products of harmonic numbers from (\ref{psi_expansion}) due to (\ref{psi_to_H}).

At this point, we will choose to concentrate on the MHV case ($\eta=1$), and come back to examine what changes for the simpler NMHV case at the end of this subsection. There exist three additional steps we need in order to bring these terms in a form where we will be able to perform the summation over all poles $k$. First of all, we can always partial fraction in order to get only inverse powers of either $k$ or $k+1$, since the degree of the numerator is always smaller than the degree of the denominator, and in particular zero. Secondly, we can perform certain manipulations in order to reduce the number of different arguments in the sums. This procedure, generally known as synchronization \cite{Vermaseren:1998uu}, is necessary for the efficient evaluation of sums of this type with the help of the computer.

In our case, achieving this goal involves treating the terms with powers of $k$ and $k+1$ differently. In particular, we can eliminate the latter in terms of the former by redefining the summation index $k^\prime=k+1$, and extending the summation range to also include $k^\prime=1$ by adding and subtracting the term in question. More concretely,
\be\label{k+1_indexshift}
\sum_{k=1}^\infty \frac{e^{-\sigma} (-e^{-2\sigma})^{k}}{(k+1)^r}\ldots=\sum_{k^\prime=2}^\infty \frac{-e^{\sigma} (-e^{-2\sigma})^{k^\prime}}{k^{\prime r}}\ldots=\left(\sum_{k=1}^\infty \frac{-e^{\sigma} (-e^{-2\sigma})^{k}}{k^{r}}\ldots\right)-\left(e^{-\sigma}\ldots\right)
\ee
and notice that this redefinition of the summation index will turn $e^{-\sigma}\to -e^{\sigma}$ in (\ref{generic_residue_term}). In addition, the argument of the harmonic numbers included in these terms will now be $k-1$, and we can turn this into the argument of the harmonic numbers multiplying inverse powers of $k$, by replacing
\be\label{shift_H}
S_{m}(k)=S_{m}(k-1)+\frac{1}{k^m}\,,
\ee
The formula above follows immediately from the definition (\ref{HarmonicN_definition}), and is the analogue of (\ref{psi_recurrence}) for harmonic numbers. Evidently we do not need to partial fraction again, as we only obtain additional powers of the same monomial in $k$.

Finally, we will make use of the fact that harmonic numbers are the simplest case of a more general set of nested sums, known as $Z$-sums \cite{Moch:2001zr}. These are defined as
\be\label{Zsum}
Z(n;m_1,\ldots,m_j;x_1,\ldots,x_j)=\sum_{n\ge i_1>i_2>\ldots>i_j>0}\frac{x_1^{i_1}}{i_1^{m_1}}\ldots\frac{x_j^{i_j}}{i_j^{m_j}}\,,
\ee
or recursively by
\be
Z(n;m_1,\ldots,m_j;x_1,\ldots,x_j)=\sum_{i_1=1}^n\frac{x_1^{i_1}}{i_1^{m_1}}Z(i_1-1;m_2,\ldots,m_j;x_2,\ldots,x_j)\,,
\ee
where $Z(n)$ is equal to the unit step function. The sum of $m_i$ is known as the weight, or transcendentality, and the number of summations $j$ as the depth.  Clearly, harmonic numbers are depth-1 $Z$-sums, $S_{m}(k)=Z(k;m;1)$.

An important property of these objects, also known as the quasi-shuffle algebra\footnote{In fact, the quasi-shuffle algebra forms part of a larger Hopf algebra structure \cite{Moch:2001zr}.}, is that a product of two $Z$-sums with the same outer summation index can be reexpressed as a linear combination of single $Z$-sums. This easily follows by splitting the square double summation range into regions with definite index ordering, namely
\be
\sum_{i=1}^n \sum_{j=1}^n a_{ij}=\sum_{i=1}^n \sum_{j=1}^{i-1} a_{ij}+\sum_{j=1}^n \sum_{i=1}^{j-1} a_{ij}+\sum_{i=1}^n a_{ii}\,.
\ee

For our purposes, rather than the general recursive procedure for decomposing a product of $Z$-sums, we will just need the particular case (we set all $x_i=1$, and for compactness drop both them and the outer summation index $n$ from our notation),
\be
\begin{aligned}\label{Z_algebra}
Z(l)Z(m_1,\ldots,m_j)&=Z(l,m_1,\ldots,m_j)+Z(m_1,l,\ldots,m_j)+\ldots+ Z(m_1,\ldots,m_j,l)\\
&\quad+ Z(m_1+l,\ldots,m_j)+\ldots+Z(m_1,\ldots,m_j+l)\,.
\end{aligned}
\ee
In other words we take all permutations that preserve the order of indices of the two $Z$-rums on the left hand side, and also add the $l$ index to all $m_i$. The above formula can then be used recursively in order to decompose the product of harmonic numbers/depth-1 $Z$-sums into single harmonic sums in (\ref{generic_residue_term}).

After these three steps, the $k$-dependent part of all terms in the sum over residues (\ref{generic_residue_term}) will be itself proportional to a $Z$-sum,
\be
\sum_{k=1}^\infty\frac{(-e^{-2\sigma})^{k}}{k^{m_1}}Z(k-1;m_2,\ldots,m_j;1,\ldots,1)=Z(\infty;m_1,m_2,\ldots,m_j;-e^{-2\sigma},1,\ldots,1)\,.
\ee
The critical observation, based on \cite{Moch:2001zr}, is that this particular $Z$-sum precisely coincides with the series representations of the harmonic polylogarithm $H_{m_1,m_2,\ldots,m_j}(-e^{-2\sigma})$ of Remiddi and Vermaseren \cite{Remiddi:1999ew}\footnote{For an earlier, implicit definition of harmonic polylogarithms in terms of the inverse Mellin transform of nested sums, see also \cite{Blumlein:1998if}.}!

We have therefore rigorously proven that the integral yielding the single-particle contribution to the Wilson loop OPE of the MHV hexagon (\ref{r_observable}) will have a weak coupling expansion of the form (\ref{hex_integral}), with its $\sigma$ dependence always given by
\be\label{h_MHV}
h^{(l)}_n(\sigma)=\sum_{s,r,m_i} c^{\pm}_{s,m_1,\ldots,m_r} e^{\pm\sigma}\sigma^s H_{m_1,\ldots,m_r}(-e^{-2\sigma})\,,\quad m_i\ge1\,,
\ee
where $c^{\pm}$ are numerical coefficients.

Finally, let us go back and extend the last part of our analysis to the case of the NMHV hexagon component integral, namely (\ref{hex_integral}), (\ref{htilde_genform1}) with $\eta=0$. We recall that since the latter formula will not contain any $\psi^{(m)}(\frac{3}{2}\pm i u)$ or inverse powers of $u\pm\frac{i}{2}$, the sum over residues (\ref{generic_residue_term}) will not contain any inverse powers of $k, k+1$. Thus we may directly reexpress the harmonic numbers as a linear combination of $Z$-sums with the help of the the quasi-algebra (\ref{Z_algebra}) in (\ref{generic_residue_term}), and obtain
\bea
\sum_{k=0}^\infty\text{Res}(\tilde h^{(l)}_n,u=\tfrac{2k+1}{2} i)&\sim&\sigma^s e^{-\sigma}\sum_{k=0}^\infty (-e^{-2\sigma})^{k} Z(m_1,\ldots,m_r)\nonumber\\
&\sim& \sigma^s e^{-\sigma}\frac{d}{dy}\left.\sum_{k=0}^\infty \frac{y^{k+1}}{k+1}Z(m_1,\ldots,m_r)\right|_{y=-e^{-2\sigma}}\\
&\sim& \sigma^s e^{-\sigma}\frac{d}{dy}\left. H_{1,m_1,\ldots,m_r}(y)\right|_{y=-e^{-2\sigma}}=\frac{\sigma^s e^{-\sigma}}{1+e^{-2\sigma}}H_{m_1,\ldots,m_r}(-e^{-2\sigma})\,.\nonumber
\eea
The equality of the last line follows from the definition of HPLs (\ref{HPL_definition}), also taking into account the discussion about their ``a''- and `m''-notation before equation (\ref{HPL_a_to_m_notation}). We similarly conclude that the weak coupling expansion of the single-particle contribution to the Wilson loop OPE of the NMHV hexagon component (\ref{NMHV_component}) will be proportional to (\ref{hex_integral}) with
\be\label{h_NMHV}
h^{(l)}_n(\sigma)=\frac{1}{2\cosh \sigma}\sum_{s,r,m_i} c_{s,m_1,\ldots,m_r} \sigma^s H_{m_1,\ldots,m_r}(-e^{-2\sigma})\,,\quad m_i\ge1\,.
\ee
The general, all-loop structure of the hexagon Wilson loop OPE contributions for the MHV (\ref{h_MHV}) and NMHV (\ref{h_NMHV}) case agree with the ans\"atze, based on empirical evidence\footnote{The fact that harmonic polylogarithms is a suitable basis for describing the hexagon Wilson loop OPE was first noted in \cite{papathanasiou_mathematica}.}, used in \cite{Basso:2013aha}\footnote{Note however that this paper uses the ``a''-notation for harmonic polylogarithms, whereas we are using the ``m''-notation. The relation between the two notations is discussed in appendix \ref{HPL_review}.}.

Before concluding, we should note that by refining our analysis so as to keep track of the maximum powers of $g$ that can multiply the terms of the integrand (\ref{htilde_genform1}), it is possible to show that the indices of the terms summing up to $h^{l}_n(\sigma)$ (\ref{h_MHV}), (\ref{h_NMHV}) are constrained as follows,
\be
s+\sum_{i} m_i\le 2(l-1)+\eta-n\,,
\ee
where again $\eta=0,1$ corresponds to the NMHV, MHV case respectively. This of course is in agreement with the fact that all $\cN=4$ amplitudes computed to date have maximal transcendentality $2L$ at $L$ loops, given also that the Taylor expansion of transcendental functions yields terms with the same or smaller transcendentality.

\section{Implementation and Results}\label{section_implementation}
\subsection{Algorithm}
In the previous section, we presented a general proof for the exact basis of harmonic polylogarithms, including the dependence of their coefficients and arguments on the kinematical data, which is suitable for describing the leading OPE contribution of the hexagon Wilson loop at any order in the weak coupling expansion $g\ll 1$.

Furthermore, it is perhaps evident that our proof forms an algorithmic process which allows us to directly compute the Fourier integrals (\ref{hex_integral}) and obtain explicit expressions for the relevant part of the hexagon in terms of the kinematical variables (\ref{crossratios}). We stress again that this method of computation can be applied at arbitrary loop order $l$, subject to restrictions in computational power.

Let us now summarize the steps of the algorithm, which facilitate its implementation in any computer algebra system. We start with the weak coupling expansion of the integrand (\ref{main_integral}), which as we reviewed in section \ref{section_WLReview}, can be performed with the help of the definitions of $\mu, \gamma, p$, and the integral (\ref{psi_integral}). Then, we
\begin{enumerate}
\item Replace all $\psi$-functions appearing with $\psi ^{(n)}\left(\tfrac{1}{2}-i \tfrac{p}{2}\right)$ with the help of (\ref{reflection_formula}) and (\ref{psi_recurrence}).
\item Replace $u=(k+\tfrac{1}{2})i+\epsilon$ and find the residue for general $k\ge 1$, with the help of (\ref{exponential_series})-(\ref{special_fraction_series}).
\item Replace $\psi^{(m-1)}(k+1)\to H_{m}(k)=Z(k,m,1)$ by means of (\ref{psi_to_H}).
\item Partial fraction the inverse powers of $k$, $k+1$, if any.
\item For the $k\ge 1$ sum over residues, synchronize the arguments by applying (\ref{k+1_indexshift}) to the terms with $k+1$ powers, and (\ref{shift_H}) to terms with $k$ powers.
\item Reduce products of $Z$-sums to linear combinations of single $Z$-sums by recursively applying (\ref{Z_algebra}).
\item Replace

\vspace{-38pt}
\[
\begin{aligned}
\sum_{k=1}^\infty\frac{(-e^{-2\sigma})^{k}}{k^{m_1}}Z(k-1;m_2,\ldots,m_j;1,\ldots,1)\to H_{m_1,m_2,\ldots,m_j}(-e^{-2\sigma})\,,\\ \sum_{k=0}^\infty(-e^{-2\sigma})^{k}Z(k;m_2,\ldots,m_j;1,\ldots,1)\to\frac{1}{1+e^{-2\sigma}} H_{m_2,\ldots,m_j}(-e^{-2\sigma})\,.
\end{aligned}
\]
\item Obtain final result for the integral by evaluating and adding $\text{Res}(h^{(l)}_n,u=\frac{i}{2})$ to the result of the previous step.
\end{enumerate}

\subsection{New predictions: MHV hexagon}
With the steps of the algorithm set in place, we can now proceed with its implementation, in order to determine the leading OPE contribution of the hexagon remainder function to high loop order. We remind the reader that this contribution will be given in terms of the $h^{(l)}_n(\sigma)$ functions as
\be
R=2\cos\phi \,e^{-\tau}\sum_{l=1}^\infty g^{2l} \left[\sum_{n=0}^{l-1}\tau^n h^{(l)}_n(\sigma)+\tfrac{\Gamma^l_\text{cusp}}{4}\left[e^{-\sigma } \log({1+e^{2 \sigma }})+ e^{\sigma } \log(1+e^{-2 \sigma })\right]
\right]+\mathcal{O}(e^{-2\tau})\,,
\ee
by virtue of equations (\ref{r_observable}),(\ref{r_to_R}),(\ref{r_BDS}),(\ref{gamma_cusp}) and (\ref{hex_integral}).

We have included the results, together with a Mathematica code that generates them, in ancillary text files accompanying the version of this paper on the arXiv. The code is fully general, and can be used in principle for any given value of $l$. In particular, we have used it to obtain $h^{(l)}_n(\sigma)$ for all allowed values of $l-1\ge n\ge0$, up to $l=6$, thereby providing information about the 5- and 6-loop hexagon for the first time.

Generating results at higher loops from this code is only a matter of computational power and optimization, and to illustrate this we also calculated the $n=l-1, l-2$ components up to $l=12$. As far as the efficiency of the code is concerned, we should stress that even without particular attention to optimization, the computation of all 4-loop leading-twist terms takes about 10 seconds on a portable computer, and all 5-loop terms about a minute!

Let us start by briefly mentioning what is already known about the hexagon in the near-collinear limit, and in the first few orders at weak coupling. In \cite{Basso:2013aha}, the $n=0$, also dubbed ``flattened'' part of the hexagon was computed up to 4 loops under the assumption of an ansatz for the general structure of the expression. More specifically, the free parameters of the ansatz were determined by comparing its Taylor expansion with a finite number of terms in the sum over residues, which the Fourier integral (\ref{hex_integral}) reduces to\footnote{See also \cite{Dixon:2012yy} for an application of the same method in the multi-Regge limit.}.

In the previous section, we proved that the structure of this ansatz is correct at any loop order, which thus places the aforementioned calculations on a firmer setting. As a further consistency check, we first aimed to reproduce the results reported in \cite{Basso:2013aha} for the $h^{(l)}_0(\sigma)$ functions\footnote{In the notations of \cite{Basso:2013aha}, $h^{(l)}_0(\sigma)\to f_l (\sigma)$.}, $n=1,\ldots,4$ with the help of our direct computation method.

More specifically, we compared with the expressions contained in the Mathematica file \texttt{Functionshf.nb} accompanying the latter paper, and found indeed agreement. Due to the many functional identities between harmonic polylogarithms, the two expressions are not identical, but they can be brought in the same form with the help of the \texttt{HPL} package \cite{Maitre:2005uu,Maitre:2007kp}. To this end, we use property (\ref{HPL_flipsign}) in order to change the argument of the HPLs to $e^{-2\sigma}$, and also replace all powers of $\sigma\to-\frac{1}{2}H_0(e^{-2\sigma})$ multiplying them. Then, we employ the command \texttt{HPLProductExpand} in both expressions, which eliminates any products of HPLs, in favor of their linear combinations.
\iffalse
\be
\mathcal{W}^{(l)}=g^{2l}\cos\phi e^{-\tau}\sum_{n=0}^{l-1}\tau^n h^{(l)}_n(\sigma)
\ee

\be
h^{(1)}_0=\frac{H_1(-x)}{\sqrt{x}}+\sqrt{x} \left(H_0(x)+H_1(-x)\right)
\ee

\be
\begin{aligned}h^{(2)}_0&=-2 \left(\sqrt{x}+\frac{1}{\sqrt{x}}\right) \left(H_0(x) H_{1,1}(-x)+2 H_{1,1}(-x)+2 H_{1,1,1}(-x)+H_0(x) H_1(-x)\right.\\
&\quad\left.+2 H_1(-x)+H_3(-x)\right)-\frac{1}{3} \left(12+\pi ^2\right) \sqrt{x} H_0(x)
\end{aligned}
\ee
\fi

Next, we moved on to determine all non-flattened leading-twist contribitions $h^{(l)}_n(\sigma)$, $n\ne 0$ for $l\le4$. As all HPLs in the expressions have argument $-e^{-2\sigma}$, we will omit it for compactness. At two loops we have
\be
h^{(2)}_1=e^\sigma \left[(4 \sigma -4)H_1-4 H_{1,1}\right]+(\sigma\to-\sigma)\,,
\ee
at three loops\footnote{Recently \cite{Dixon:2013eka} appeared, which calculates the 3-loop hexagon in general kinematics, and also specializes to the near-collinear kinematics considered here. We have checked that the results of the two calculations agree if we take into account that the expansion parameter in the latter reference is $2g^2$. See also \cite{Bartels:2011xy} for a calculation of the analogue of $h^{(3)}_2$ in the earlier Wilson loop OPE approach \cite{Alday:2010ku}.}
\begin{align}
h^{(3)}_2=&e^\sigma \left[-16 (\sigma -1) H_{1,1}+16 H_{1,1,1}+H_1 \left(4 \left(\sigma ^2-4 \sigma +3\right)+\tfrac{\pi ^2}{3}\right)-2H_2 (2 \sigma +1)-4 H_3\right]\nonumber\\
&+(\sigma\to-\sigma)\,,
\end{align}

\vspace{-25pt}
\begin{align}
\vspace{20pt}h^{(3)}_1=&e^\sigma \left[H_1 \left(-4 \zeta _3+8 \sigma ^2-\tfrac{8 \pi ^2 \sigma }{3}-40 \sigma +2 \pi ^2+36\right)-2H_2 (2 \sigma +1)+H_3 (2-4 \sigma )-8 \sigma  H_{1,2}\right.\nonumber\\
&+2 \left(20+\pi ^2-24 \sigma +4 \sigma ^2\right) H_{1,1}-4(2 \sigma +1) H_{2,1}+48(1- \sigma ) H_{1,1,1}+48 H_{1,1,1,1}\Big].\nonumber\\
&+(\sigma\to-\sigma)\,,
\end{align}
whereas at four loops
\begin{align}
h^{(4)}_3=&e^\sigma \left[H_1 \left(8 \zeta _3+\tfrac{16 \sigma ^3}{9}-16 \sigma ^2+\tfrac{4 \pi ^2 \sigma }{9}+48 \sigma -\tfrac{4 \pi ^2}{3}-\tfrac{80}{3}\right)+H_2 \left(-\tfrac{16 \sigma ^2}{3}+8 \sigma -\tfrac{4 \pi ^2}{9}+\tfrac{32}{3}\right)\right.\nonumber\\
&+H_3 \left(\tfrac{32}{3}-\tfrac{16 \sigma }{3}\right)+\left(-16 \sigma ^2+64 \sigma -\tfrac{4 \pi ^2}{3}-48\right) H_{1,1}+16 \sigma  H_{1,2}+(16 \sigma +8) H_{2,1}\nonumber\\
&\left.+(64 \sigma -64) H_{1,1,1}+\tfrac{40 H_{1,3}}{3}+8 H_{2,2}+16 H_{3,1}-64 H_{1,1,1,1}\right]+(\sigma\to-\sigma)\,,
\end{align}
\begin{align}
h^{(4)}_2=&e^\sigma \left[H_1 \left(-8 \zeta _3 \sigma +40 \zeta _3+\tfrac{16 \sigma ^3}{3}-\tfrac{16 \pi ^2 \sigma ^2}{3}-64 \sigma ^2+\tfrac{52 \pi ^2 \sigma }{3}+224 \sigma -\tfrac{7 \pi ^4}{15}-\tfrac{40 \pi ^2}{3}-160\right)\right.\nonumber\\
&+H_2  \left(8 \zeta _3-8 \sigma ^2+4 \pi ^2 \sigma +24 \sigma +\tfrac{2 \pi ^2}{3}+32\right)+H_3  \left(8 \sigma +\tfrac{8 \pi ^2}{3}+16\right)+H_4 (16 \sigma +8)\nonumber\\
&+24 H_5+H_{1,1} \left(40 \zeta _3+\tfrac{16 \sigma ^3}{3}-80 \sigma ^2+\tfrac{52 \pi ^2 \sigma }{3}+304 \sigma -\tfrac{52 \pi ^2}{3}-224\right)+(8 \sigma +40) H_{1,3}\nonumber\\
&+H_{1,2} \left(-16 \sigma ^2+80 \sigma -\tfrac{4 \pi ^2}{3}\right)+H_{2,1} \left(-16 \sigma ^2+40 \sigma -\tfrac{4 \pi ^2}{3}+40\right)+(24 \sigma +24) H_{2,2}\nonumber\\
&+H_{1,1,1} \left(-80 \sigma ^2+384 \sigma -\tfrac{52 \pi ^2}{3}-304\right)+80 \sigma  (H_{1,1,2}+  H_{1,2,1})+(80 \sigma +40) H_{2,1,1}\nonumber\\
&+384( \sigma -1) H_{1,1,1,1}+8 H_{1,4}+16 (H_{2,3}+H_{3,2})+24 H_{3,1}+40 (H_{1,1,3}+H_{1,3,1})\nonumber\\
&+24 (H_{1,2,2}+ H_{2,1,2}+ H_{2,2,1})+48 H_{3,1,1}-384 H_{1,1,1,1,1}\Big]+(\sigma\to-\sigma)\,,
\end{align}
\be
\begin{aligned}
h^{(4)}_1=&e^{\sigma}\left[\left(\tfrac{4 \pi ^2 \sigma ^3}{9}+8 \sigma ^3-8 \zeta _3 \sigma ^2-12 \pi ^2 \sigma ^2-112 \sigma ^2+\tfrac{113 \pi ^4 \sigma }{45}+\tfrac{122 \pi ^2 \sigma }{3}+464 \sigma +\tfrac{8 \pi ^2 \zeta _3}{3}+76 \zeta _3\right.\right.\\
&\,\left.+32 \zeta _5-\tfrac{91 \pi ^4}{45}-28 \pi ^2-400\right)H_1+\left(\tfrac{4 \pi ^2 \sigma }{3}+20 \sigma +2 \pi ^2\right) H_3+\left(8 \sigma ^2+4 \sigma +\tfrac{2 \pi ^2}{3}-8\right) H_4\\
&+H_2 \left(-\tfrac{4}{3} \pi ^2 \sigma ^2-8 \sigma ^2+\tfrac{14 \pi ^2 \sigma }{3}+48 \sigma +12 \zeta _3-\tfrac{\pi ^4}{9}+\tfrac{8 \pi ^2}{3}+32\right)+(24 \sigma -24) H_5\\
&+\left(\tfrac{32 \sigma ^3}{3}-12 \pi ^2 \sigma ^2-152 \sigma ^2+\tfrac{152 \pi ^2 \sigma }{3}+592 \sigma +96 \zeta _3-\tfrac{91 \pi ^4}{45}-38 \pi ^2-464\right) H_{1,1}+8 \sigma  H_{1,4}\\
&+\left(\tfrac{8 \sigma ^3}{3}-40 \sigma ^2+10 \pi ^2 \sigma +152 \sigma +20 \zeta _3-\tfrac{10 \pi ^2}{3}\right) H_{1,2}+\left(16 \sigma +\tfrac{10 \pi ^2}{3}+60\right) H_{1,3}-20 H_{1,5}\\
&+\left(-16 \sigma ^2+\tfrac{28 \pi ^2 \sigma }{3}+64 \sigma +24 \zeta _3+2 \pi ^2+64\right) H_{2,1}+\left(-8 \sigma ^2+36 \sigma +\tfrac{4 \pi ^2}{3}+32\right) H_{2,2}\\
&+(8 \sigma +12) (H_{2,3}+H_{3,2})-12 (H_{2,4}+H_{4,2}+H_{3,3})+\left(24 \sigma +4 \pi ^2+20\right) H_{3,1}-24 H_{5,1}\\
&+(8 \sigma +4) H_{4,1}+\left(\tfrac{32 \sigma ^3}{3}-192 \sigma ^2+\tfrac{152 \pi ^2 \sigma }{3}+768 \sigma +96 \zeta _3-48 \pi ^2-592\right) H_{1,1,1}\\
&+\left(-40 \sigma ^2+192 \sigma -\tfrac{10 \pi ^2}{3}\right) (H_{1,1,2}+H_{1,2,1})+(16 \sigma +80) (H_{1,1,3}+H_{1,3,1})+96 H_{3,1,1,1}\\
&+(56 \sigma +40) (H_{1,2,2}+H_{2,1,2})+20 (H_{1,2,3}+H_{1,3,2})+\left(-32 \sigma ^2+96 \sigma -\tfrac{8 \pi ^2}{3}+88\right) H_{2,1,1}\\
&+24 (H_{2,1,3}+H_{2,3,1}H_{2,2,2}+H_{3,1,2}+ H_{3,2,1})+(56 \sigma +44) H_{2,2,1}+96(1+2 \sigma) H_{2,1,1,1}\\
&+\left(-192 \sigma ^2+960 \sigma -48 \pi ^2-768\right) H_{1,1,1,1}+192 \sigma  (H_{1,1,1,2}+H_{1,1,2,1}+H_{1,2,1,1})\\
&+80 (H_{1,1,1,3}+H_{1,1,3,1}+H_{1,3,1,1})+40 (H_{1,1,2,2}+ H_{1,2,1,2}+ H_{1,2,2,1})+(16 \sigma +40) H_{3,1,1}\\
&+48 (H_{2,1,1,2}+ H_{2,1,2,1}+ H_{2,2,1,1})+960 (\sigma -1) H_{1,1,1,1,1}-960 H_{1,1,1,1,1,1}\Big]+(\sigma\to-\sigma)\,.
\end{aligned}
\ee
Evidently the length of the expressions grows quite fast with the loop order, due to the increase not only in the number of terms in the integrand, but also in the quasi-shuffle algebra decomposition (\ref{Z_algebra}) for higher powers of harmonic numbers.

\begin{figure}
\centering
\includegraphics[width=0.6\textwidth]{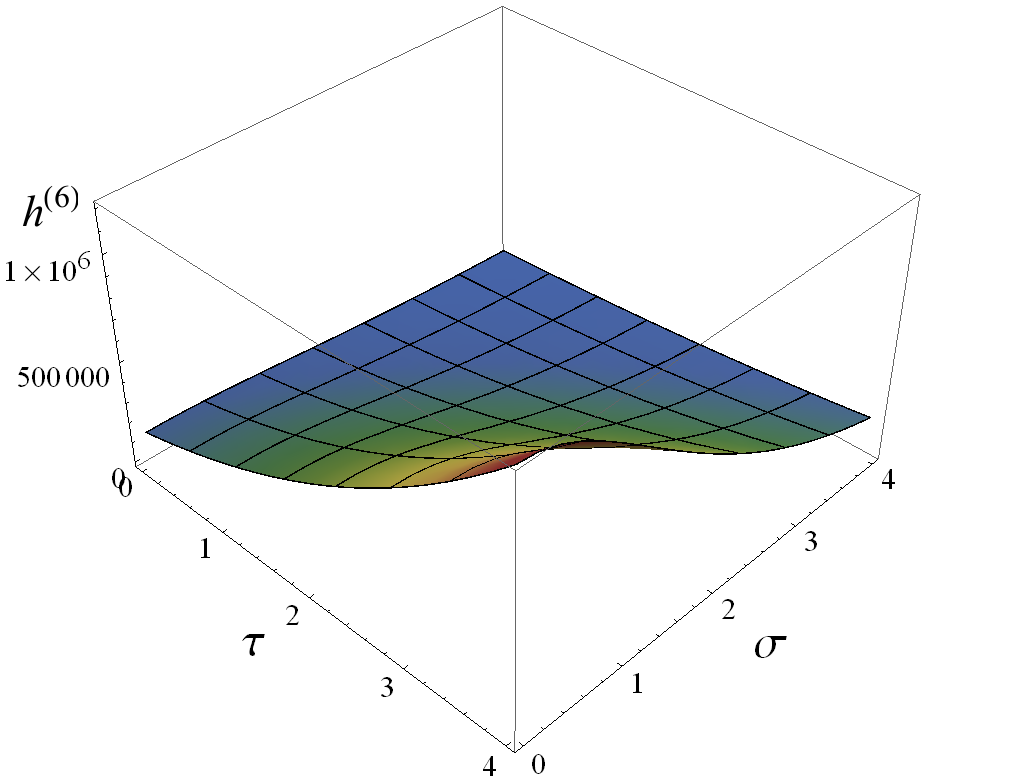}
\caption{Plot of the MHV hexagon leading OPE contribution at 6 loops, $h^{(6)}\equiv\sum_{n=0}^{5}\tau^n h^{(6)}_n(\sigma)$, as a function of $\tau, \sigma$. Colors of the visible spectrum denote different values of $h^{(6)}$, increasing from blue to red. The function is always positive, and monotonically increasing and decreasing in $\tau$ and $\sigma$ respectively.}
\label{fig:3dMHVplot}
\end{figure}

\begin{figure}
\centering
\includegraphics[width=0.94\textwidth]{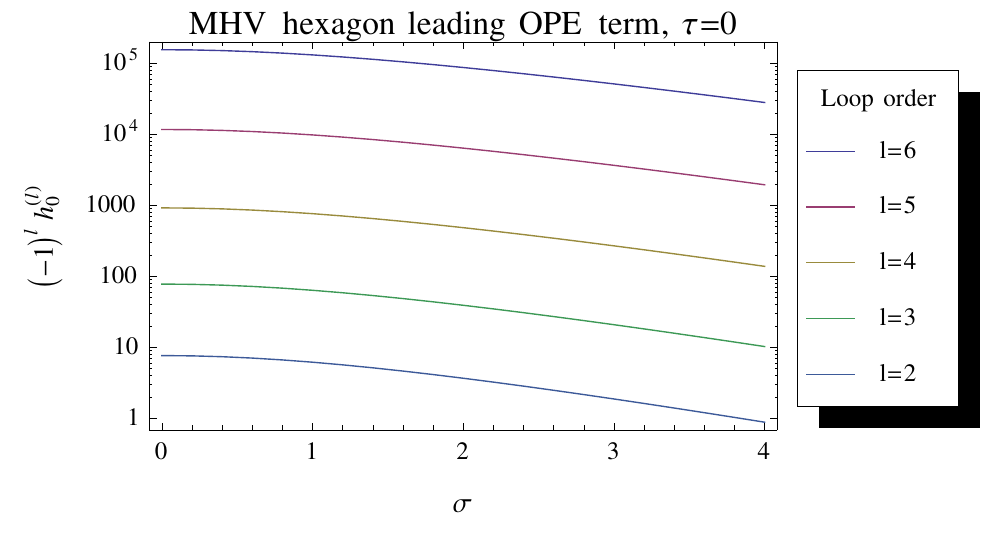}
\caption{Log-linear plot of the $h^{(l)}_0$ component, yielding the leading OPE contribution at $\tau=0$, as a function of $\sigma$ at different loop orders $l$. Its sign is given by $(-1)^l$, as it varies continuously without vanishing. Increasing $l$ by one increases the magnitude by roughly a factor of 10, maintaining similar shape.}
\label{fig:2DMHVplot}
\end{figure}

Finally, we computed all 5- and 6-loop leading-twist contributions to the OPE of the MHV hexagon, as well as the $h^{(l)}_{l-1}(\sigma)$ and $h^{(l)}_{l-2}(\sigma)$ terms up to $l=12$ loops. For compactness, we will content ourselves with writing down only the simplest 5-loop term,
\be
\begin{aligned}
h^{(5)}_4=&e^\sigma \left[H_1 \left(\tfrac{32 \zeta _3 \sigma }{3}-32 \zeta _3+\tfrac{4 \sigma ^4}{9}-\tfrac{64 \sigma ^3}{9}+\tfrac{2 \pi ^2 \sigma ^2}{9}+48 \sigma ^2-\tfrac{16 \pi ^2 \sigma }{9}-\tfrac{320 \sigma }{3}+\tfrac{7 \pi ^4}{540}+4 \pi ^2+\tfrac{140}{3}\right)\right.\\
&+H_2 \left(-\tfrac{32 \zeta _3}{3}-\tfrac{8 \sigma ^3}{3}+\tfrac{40 \sigma ^2}{3}-\tfrac{2 \pi ^2 \sigma }{3}-\tfrac{16 \sigma }{3}+\tfrac{10 \pi ^2}{9}-30\right)+H_3 \left(-\tfrac{4 \sigma ^2}{3}+16 \sigma -\tfrac{\pi ^2}{9}-\tfrac{58}{3}\right)\\
&+H_4 (4 \sigma +2)+4 H_5+H_{1,1} \left(-32 \zeta _3-\tfrac{64 \sigma ^3}{9}+64 \sigma ^2-\tfrac{16 \pi ^2 \sigma }{9}-192 \sigma +\tfrac{16 \pi ^2}{3}+\tfrac{320}{3}\right)\\
&+\left(\tfrac{64 \sigma ^2}{3}-64 \sigma +\tfrac{16 \pi ^2}{9}\right) H_{1,2}+\left(\tfrac{64 \sigma ^2}{3}-32 \sigma +\tfrac{16 \pi ^2}{9}-\tfrac{128}{3}\right) H_{2,1}+\left(\tfrac{32 \sigma }{3}-\tfrac{160}{3}\right) H_{1,3}\\
&+\left(64 \sigma ^2-256 \sigma +\tfrac{16 \pi ^2}{3}+192\right) H_{1,1,1}+\left(-\tfrac{32 \sigma }{3}-32\right) H_{2,2}+\left(\tfrac{64 \sigma }{3}-\tfrac{128}{3}\right) H_{3,1}-\tfrac{32}{3}H_{1,4}\\
&-64 \sigma  (H_{1,1,2}+  H_{1,2,1})-32(2 \sigma+1) H_{2,1,1}+256(1- \sigma ) H_{1,1,1,1}-\tfrac{64}{3}(H_{2,3}+ H_{3,2})\\
&-\tfrac{160}{3} (H_{1,1,3}+H_{1,3,1})-32 (H_{2,1,2}+ H_{2,2,1}+H_{1,2,2})-64 H_{3,1,1}+256 H_{1,1,1,1,1}\Big]\\
&+(\sigma\to-\sigma)\,.
\end{aligned}
\ee
We also present a plot of the full 6-loop $\mathcal{O}(e^{-\tau})$ term for the MHV hexagon in figure \ref{fig:3dMHVplot}, as well as a comparison of the $h^{(l)}_0(\sigma)$ component at different loop orders in figure \ref{fig:2DMHVplot}.

As mentioned in the introductory remarks of this subsection, the rest of our results at 5 loops and beyond are included in the ancillary files accompanying this article. In more detail, the Mathematica implementation of the algorithm is included in \texttt{algorithm.m}, the full single-particle OPE contribution up to 6 loops in \texttt{MHV\_full.m}, and the part of the latter with leading and subleading powers of $\tau$ up to 12 loops in \texttt{MHV\_leadtau.m} and \texttt{MHV\_subleadtau.m} respectively.

\subsection{New predictions: NMHV hexagon}
Similarly, we employed our algorithm implementation in order to compute all $\mathcal{O}(e^{-\tau})$ contributions in the OPE of the NMHV hexagon component (\ref{NMHV_component}) up to 6 loops, and all of its leading and subleading powers of $\tau$ up to 12 loops. Also in this case, our results agreed with the previous indirect computation of \cite{Basso:2013aha}, contained in the file \texttt{Fnl.nb} attached to the latter paper, up to 3 loops.

Of our new results, for reasons of space we will only mention here three out of the five single-particle contributions at 4 loops, saving the rest for the attached ancillary files. After we rescale the overall $\sigma$-dependent factor in (\ref{h_NMHV}), and also shift the loop index so as to measure the total powers of $g^2$ in (\ref{NMHV_component}),
\be\label{h_NMHV_rescale}
h^{(l+1)}_n(\sigma)=\frac{1}{2\cosh \sigma}F^{(l)}_n(\sigma)\,,
\ee
the aforementioned component of the NMHV ratio function will be given by
\be
\mathcal{R}^{(6134)}= \frac{e^{-\tau}}{2\cosh\sigma}\sum_{l=0}^\infty g^{2l} \sum_{n=0}^{l}\tau^n F^{(l)}_n(\sigma)+\mathcal{O}(e^{-2\tau})\,,\label{NMHVratio_final}
\ee
where
\begin{align}
F^{(4)}_4=&-\tfrac{16}{9} H_1 \left(18 \zeta _3+4 \sigma ^3+\pi ^2 \sigma \right)+\tfrac{16}{9} H_2 \left(12 \sigma ^2+\pi ^2\right)+\tfrac{32 \sigma }{3}H_3-\tfrac{32 }{3}H_4-64 \sigma  (H_{1,2}+  H_{2,1})\nonumber\\
&+\tfrac{16}{3} \left(12 \sigma ^2+\pi ^2\right) H_{1,1}-256 \sigma  H_{1,1,1}-32 H_{2,2}-\tfrac{160 }{3}(H_{1,3}+H_{3,1})+256 H_{1,1,1,1}\nonumber\\
&+\tfrac{4}{9} \left(24 \zeta _3 \sigma +\sigma ^4\right)+\tfrac{2 \pi ^2 \sigma ^2}{9}+\tfrac{7 \pi ^4}{540}\,,
\end{align}
\begin{align}
F^{(4)}_3=&H_1 \left(\tfrac{224 \zeta _3 \sigma }{3}+\tfrac{16 \sigma ^4}{9}+\tfrac{248 \pi ^2 \sigma ^2}{9}+\tfrac{319 \pi ^4}{135}\right)+H_2 \left(-\tfrac{224 \zeta _3}{3}-\tfrac{32 \sigma ^3}{3}-24 \pi ^2 \sigma \right)-\tfrac{64  \sigma }{3}H_4-56 H_5\nonumber\\
&+H_3 \left(\tfrac{32 \sigma ^2}{3}-\tfrac{112 \pi ^2}{9}\right)+H_{1,1} \left(-288 \zeta _3-\tfrac{448 \sigma ^3}{9}-\tfrac{880 \pi ^2 \sigma }{9}\right)+\left(\tfrac{448 \sigma ^2}{3}+\tfrac{112 \pi ^2}{9}\right)( H_{1,2} +H_{2,1})\nonumber\\
&+\left(576 \sigma ^2+112 \pi ^2\right) H_{1,1,1}+\tfrac{32}{3} \sigma  (H_{1,3}+H_{3,1})-576 \sigma  (H_{1,1,2}+ H_{1,2,1}+ \sigma  H_{2,1,1})\nonumber\\
&-\tfrac{416}{3} \sigma  H_{2,2}-2560 \sigma  H_{1,1,1,1}-\tfrac{448 }{3}(H_{2,3}+H_{3,2})-\tfrac{224 }{3}(H_{1,4}+H_{4,1})-\tfrac{1120}{3} H_{1,1,3}\nonumber\\
&-224 (H_{1,2,2}+H_{2,1,2}+H_{2,2,1})-\tfrac{1120}{3} (H_{1,3,1}+ H_{3,1,1})+2560 H_{1,1,1,1,1}\nonumber\\
&-\tfrac{16 \zeta _3 \sigma ^2}{3}-\tfrac{76 \pi ^2 \zeta _3}{9}-\tfrac{248 \zeta _5}{3}-\tfrac{32}{9} \pi ^2 \sigma ^3-\tfrac{44 \pi ^4 \sigma }{45}\,,
\end{align}
\begin{align}
F^{(4)}_2=&H_1 \left(16 \zeta _3 \sigma ^2-\tfrac{124 \pi ^2 \zeta _3}{3}-376 \zeta _5-\tfrac{112}{9} \pi ^2 \sigma ^3-\tfrac{184 \pi ^4 \sigma }{9}\right)+H_2 \left(48 \zeta _3 \sigma +\tfrac{4 \sigma ^4}{3}+\tfrac{98 \pi ^2 \sigma ^2}{3}+\tfrac{503 \pi ^4}{180}\right)\nonumber\\
&+H_3 \left(-16 \zeta _3-\tfrac{8 \pi ^2 \sigma }{3}\right)-64 H_5 \sigma -\tfrac{32 \pi ^2 H_4}{3}-42 H_6\nonumber+\left(168 \sigma ^2-18 \pi ^2\right) H_{2,2}-136 H_{3,3}\nonumber\\
&+(H_{1,2}+H_{2,1}) \left(-304 \zeta _3-\tfrac{128 \sigma ^3}{3}-\tfrac{368 \pi ^2 \sigma }{3}\right)+H_{1,1,1} \left(-1248 \zeta _3-192 \sigma ^3-560 \pi ^2 \sigma \right)\nonumber\\
&+H_{1,1} \left(224 \zeta _3 \sigma +\tfrac{16 \sigma ^4}{3}+\tfrac{424 \pi ^2 \sigma ^2}{3}+\tfrac{979 \pi ^4}{45}\right)+\left(32 \sigma ^2-\tfrac{152 \pi ^2}{3}\right) (H_{1,3}+ H_{3,1})\nonumber\\
&+\left(608 \sigma ^2+\tfrac{152 \pi ^2}{3}\right) (H_{1,1,2}+ H_{1,2,1}+ H_{2,1,1})+\left(2496 \sigma ^2+592 \pi ^2\right) H_{1,1,1,1}\nonumber\\
&-96 \sigma  (H_{1,4}+H_{4,1})-64 \sigma  (H_{2,3}+  H_{3,2})-32 \sigma  (H_{1,1,3}+H_{1,3,1}+H_{3,1,1})\nonumber\\
&-640 \sigma  (H_{1,2,2}+  H_{2,1,2}+  H_{2,2,1})-2496 \sigma  (H_{1,1,1,2}+  H_{1,1,2,1}+ H_{1,2,1,1}+H_{2,1,1,1})\nonumber\\
&-11520 \sigma  H_{1,1,1,1,1}-88 (H_{1,5}+H_{5,1})-124 (H_{2,4}+H_{4,2})\nonumber\\
&-224 (H_{1,1,4}+H_{1,4,1}+H_{4,1,1})-528 (H_{1,2,3}+H_{1,3,2}+H_{2,1,3}+H_{2,3,1}+H_{3,1,2}+H_{3,2,1})\nonumber\\\
&-1440 (H_{1,1,1,3}+H_{1,1,3,1}+H_{1,3,1,1}+H_{3,1,1,1})-528 H_{2,2,2}\nonumber\\
&-832 (H_{1,1,2,2}+ H_{1,2,1,2}+ H_{1,2,2,1}+ H_{2,1,1,2}+H_{2,1,2,1}+ H_{2,2,1,1})+11520 H_{1,1,1,1,1,1}\nonumber\\
&-\tfrac{16 \zeta _3 \sigma ^3}{3}+12 \pi ^2 \zeta _3 \sigma +64 \zeta _5 \sigma -4 \zeta _3^2+\tfrac{2 \pi ^2 \sigma ^4}{9}+\tfrac{301 \pi ^4 \sigma ^2}{45}+\tfrac{101 \pi ^6}{168}\,.
\end{align}
Notice in particular the symmetry under permutations of all HPL integer indices, as a result of ordering the independent summation ranges of the harmonic numbers in (\ref{generic_residue_term}) in all possible ways, so at to express them in terms of HPLs. In the MHV case the first HPL index is special due to the presence of inverse powers of $k, k-1$, however the permutation symmetry still holds for the remaining indices.

\begin{figure}
\centering
\begin{subfigure}[b]{0.49\textwidth}
\includegraphics[width=\textwidth]{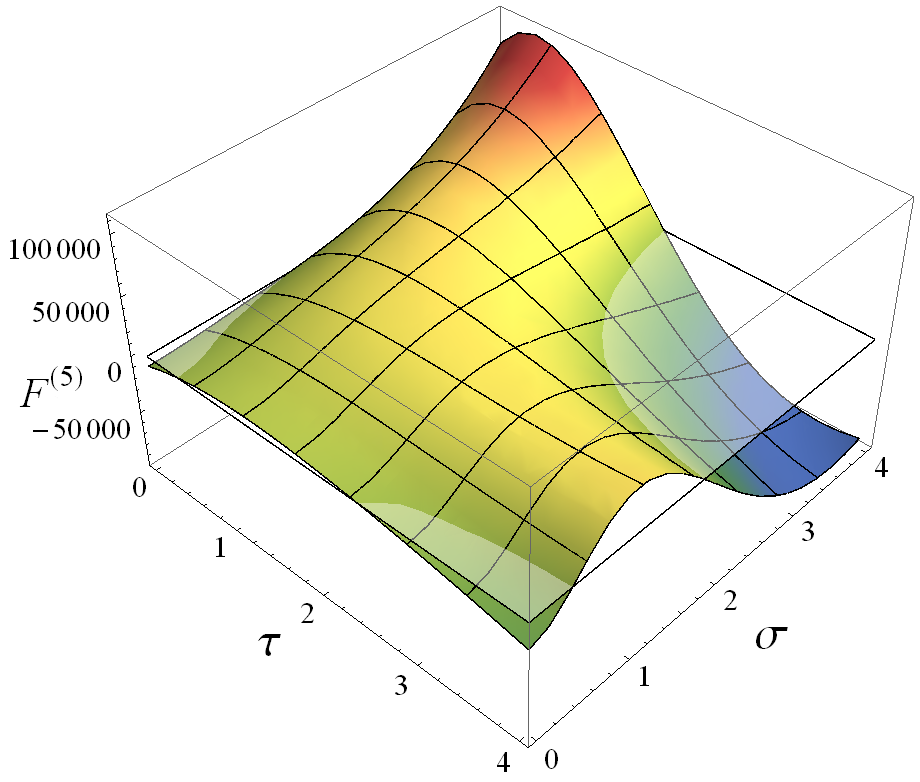}
\caption{\label{NMHVplotA}}
\end{subfigure}
\begin{subfigure}[b]{0.49 \textwidth}
\includegraphics[width=\textwidth]{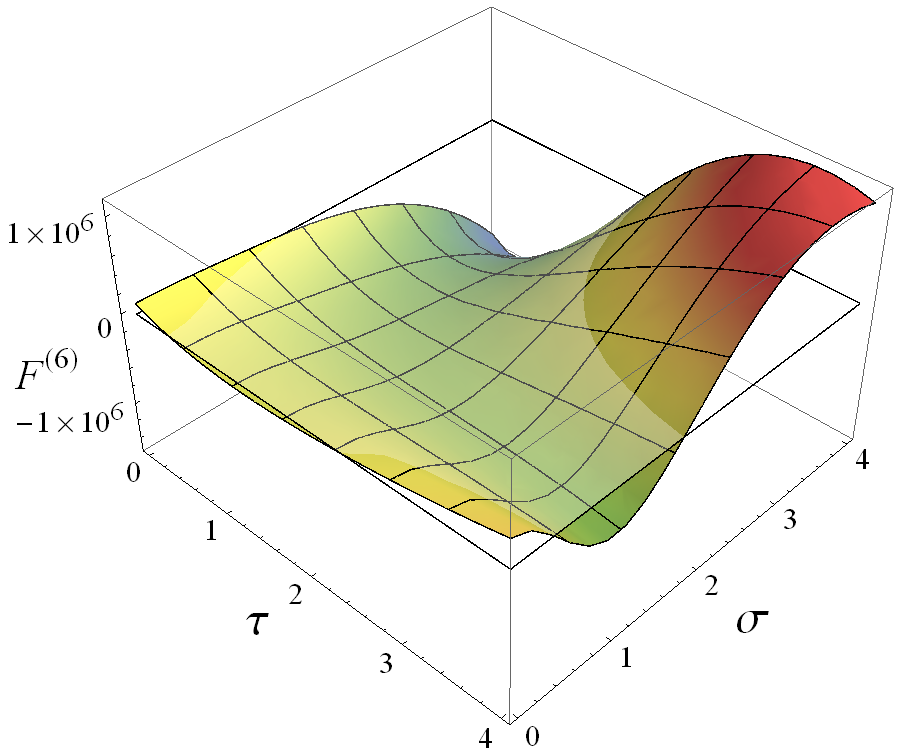}
\caption{\label{NMHVplotB}}
\end{subfigure}
\caption{Plots of the leading OPE contribution $F^{(l)}\equiv\sum_{n=0}^l \tau^n F^{(l)}_n(\sigma)$ to the NMHV ratio function component (\ref{NMHVratio_final}) at $l=5, 6$ loops, as a function of $\tau, \sigma$. As the functions change sign, we have also included the $F^{(l)}=0$ plane for comparison.}
\label{fig:NMHVplots}
\end{figure}

Plots of the full $5-$ and $6-$loop $\cO(e^{-\tau})$ OPE contribution to the NMHV ratio function component are presented in figure \ref{fig:NMHVplots}. Analytic expressions for all $F^{(l)}_n$ terms up to $l=6$ loops may be found in the file \texttt{NMHV\_full.m}, and the $F^{(l)}_l$ and $F^{(l)}_{l-1}$ for $7\le l\le12$ may be found in \texttt{NMHV\_leadtau.m} and \texttt{NMHV\_subleadtau.m} respectively.

\section{Conclusions and Outlook}
In this paper, we explored the implications of the recent conjecture \cite{Basso:2013vsa,Basso:2013aha}, which adds significant new ingredients to the OPE approach to null polygonal Wilson loops, for the case of the MHV and NMHV hexagon. We found that the integrability underlying this proposal, and manifesting itself in the presence of polygamma functions in the energy and higher conserved charges of the flux tube\footnote{In \cite{Tarasov:1983cj}, the hamiltonian density of the Heisenberg XXX, or $SL(2)$, spin chain of spin $s$ was found to be a digamma function of the operator measuring the total spin of the adjacent sites. A similar expression for the hamiltonian density of the $PSU(2,2|4)$ spin chain, giving rise to the complete 1-loop dilatation operator of $\mathcal{N}=4$ SYM, was found in \cite{Beisert:2003jj} by appropriate identification of an $SL(2)$ subsector. The generalization to the all-loop dilatation operator relies on a deformation of the spectral parameter, which loosely speaking is responsible for the appearance of polygamma functions.} , heavily constrains the conformally invariant functions parametrizing the hexagon: It implies that the leading OPE contribution for these functions is expressed in terms of harmonic polylogarithms with specific kinematical dependence, at any loop order in the weak coupling expansion!

This general result is consistent with the evidence based on a particular ``$d\log$'' form for the all-loop integrand, and presented in \cite{ArkaniHamed:2012nw}, that a basis of multiple (or Goncharov) polylogarithms is sufficient for describing the MHV and NMHV hexagon. In particular, harmonic polylogarithms are a single-variable subset of multiple polylogarithms. The basis of harmonic polylogarithms we found also agrees with the ans\"atze assumed in \cite{Basso:2013aha} in order to compute (parts of) the leading OPE contribution up to 3 loops for the NMHV case, and up to 4 for the MHV, thus further supporting their validity.

Moreover, our analysis has important consequences in the practical front of extracting data for the hexagon at higher orders in the weak coupling expansion. Starting from the Fourier integral form of the aforementioned OPE contribution as predicted by \cite{Basso:2013aha}, our proof of general structure forms an algorithm for the direct computation of these integrals, in principle at any loop order. By implementing this algorithm in Mathematica, we were able to obtain new results for the full $\mathcal{O}(e^{-\tau})$ term in the near-collinear limit of the MHV remainder function and NMHV ratio function up to 6 loops, and the part of this term with the leading and subleading powers of $\tau$ up to 12 loops. We include these results in the attached ancillary files, together with an implementation of the algorithm, which can be readily used to evaluate higher-order contributions as well.

Input from the Wilson loop OPE has been crucial, together with other reasonable assumptions, for the recent determination of the full 3-loop MHV remainder function in general kinematics \cite{Dixon:2013eka}. The authors of the latter paper report that the same methodology may be used for obtaining the remainder function at 4 loops and higher, and we hope that the data presented in this paper will again prove useful in that respect. The same applies for the NMHV ratio function, which is currently known to 2 loops \cite{Dixon:2011nj}. Furthermore, it has been argued in \cite{Dixon:2013eka}, that the hexagon MHV remainder and NMHV ratio functions are described by hexagon functions, a more restricted class of combinations of multiple polylogarithms with proper branch cuts. It would be an interesting consistency check to investigate whether their near-collinear limit reduces to the basis (\ref{h_MHV}) at arbitrary loop order.

There is a number of exciting open questions which require further inquiry. Reconstruction of the full hexagon Wilson loop from its OPE would require obtaining all terms with higher powers of $e^{-\tau}$ in the near-collinear limit, which are contributions of multiparticle states propagating in the flux tube. Indeed, it is possible to extend the current framework in order to incorporate 2-particle excitations \cite{Basso:2014koa} and higher, and one would ideally like to find a general basis/method to calculate the resulting integrals for these terms as well. Note that even though the dimensionality of the integrals will be equal to the excitation number, data from the 3-loop hexagon remainder function show that the $e^{-2\tau}$ terms are described by harmonic polylogarithms as well \cite{Dixon:2013eka}. Na\"ively similar multidimensional integrals, which are however not expected to only yield HPLs, appear in the near-collinear limit of the heptagon. For the latter, very little is known beyond the 2-loop total differential \cite{CaronHuot:2011ky,Golden:2013lha} and motivic avatar structure \cite{Golden:2013xva} for the MHV case, and symbol for the NMHV case \cite{CaronHuot:2011kk}.

Aside the near-collinear limit, a similar all-loop integral formula for the hexagon remainder function also exists in the multi-Regge limit \cite{Lipatov:2010ad,Fadin:2011we}. Based on evidence that the relevant class of functions for describing this limit are certain combinations of HPLs that are single-valued on the complex plane, the authors of \cite{Dixon:2012yy} were able to extract the (next-to-)leading-logarithmic part of the aforementioned formula to high loop order. Given the resemblance of the integral formulas in the near-collinear and multi-Regge limit, it should be possible to extend our method in order to rigorously prove the appropriateness of the single-valued HPLs, and possibly the conjecture \cite{Pennington:2012zj} generalizing the results of \cite{Dixon:2012yy} for the leading-lorarithmic part to all loops. More broadly, the similarities between the two limits seem to be quite extensive, and suggest that a physical picture based on integrability should exist for the multi-Regge limit as well. For example, the main quantities entering the master formula of \cite{Lipatov:2010ad,Fadin:2011we} only depend on polygamma functions up to the currently known order.

Finally, our work may prove useful for the computation of other observables of $\mathcal{N}=4$ super Yang-Mills theory. In \cite{Belitsky:2013xxa,Belitsky:2013bja}, a very interesting class of infrared-finite weighted cross-sections was introduced and extensively studied. These measure the total flow of charge registered by detectors positioned at spatial infinity, for an initial state generated by the scalar half-BPS operator acting on the vacuum. For the choice of charges considered in the latter papers, and for two detectors measuring them at different directions simultaneously, these were shown to be related to 4-point correlation functions of components of the stress-tensor multiplet. In particular, they are given by the convolution of the Mellin transform of the Euclidean correlators with a coupling-independent `detector kernel'. The resulting inverse Mellin integral was computed from its known ingredients at one loop, and a generalization of our method may be applicable for performing the computation at higher loops, given the relation between Fourier and Mellin transforms,
\be
\frac{1}{2\pi}\int_{-\infty}^{+\infty}dx e^{ipx}f(p)=\frac{1}{2\pi i}\int_{-i\infty}^{+i\infty}ds (e^{-x})^{-s}f(-is)\,.
\ee
We expect that again harmonic polylogarithms will play a role in describing the dependence of these double flow correlations on the single variable parametrizing them, the angle between the two detectors.

\section*{Acknowledgments}
We'd like to thank Marcus Spradlin, Christian Vergu and Anastasia Volovich for collaboration at the early stages of this project. We are also grateful to Benjamin Basso, James Drummond and Emery Sokatchev for enlightening discussions and comments on the manuscript. This work was supported in
part by the US Department of Energy under Grant No. DE-FG02-97ER-41029 at the University of Florida, and the French
National Agency for Research (ANR) under contract StrongInt (BLANC-SIMI-4-2011) at LAPTh.

\appendix

\section{Review of Useful Functions}
\subsection{The $\gamma^{\varnothing} _\pm$ functions}\label{appx_gammao_functions}
For completeness, we review the functions $\gamma^{\varnothing} _\pm$ \cite{Basso:2010in} which enter in the expressions (\ref{Ep_gluon}) and (\ref{Ep_scalar}), for the anomalous dimensions and momenta of the flux tube excitations propagating between the two parts of the polygon.

These are defined as
\be\label{gamma_vacpm}
\begin{aligned}
\gamma^{\varnothing}_{-}(t) &= 2\sum_{n=1}^\infty(2n-1)\gamma^{\varnothing}_{2n-1}J_{2n-1}(t)\, , \\
\gamma^{\varnothing}_{+}(t) &= 2\sum_{n=1}^\infty(2n)\gamma^{\varnothing}_{2n}J_{2n-1}(t)\, ,
\end{aligned}
\ee
where the coefficients $\gamma^{\varnothing}_{n}$ depend on $g$ and obey
\be
\gamma^{\varnothing}_{n} + \int_{0}^{\infty}{dt\over t}J_{n}(2gt){\gamma^{\varnothing}_{+}(2gt)-(-1)^n\gamma^{\varnothing}_{-}(2gt) \over e^{t}-1} = 2g\, \delta_{n, 1}\,.
\ee
The last two equations can be solved perturbatively in $g\ll1$, with the help of the following Taylor explansion of the $i$-th Bessel function of the first kind $J_i(z)$ around $z=0$,
\be\label{BesselJ_expansion}
J_i(z)=\sum _{n=0}^{\infty } \frac{(-1)^n }{k! \Gamma (i+n+1)}\left(\frac{z}{2}\right)^{i+2 n}\,,
\ee
In any case, from the latter formula it is evident that for small $t$, $\gamma^{\varnothing}_\pm$ have a regular Taylor expansion.

\subsection{The $f_i$ functions}\label{appx_f_functions}
In this appendix, we review the computation of the $f_i$ functions \cite{Basso:2013aha} appearing in the integration measure (\ref{measure_gluon}), (\ref{measure_scalar}) of the leading (N)MHV hexagon OPE contribution.

The functions in question are defined as
\be\label{f_definitions}
\begin{aligned}
f_1(u,v)&=2 \, \tilde \kappa(u) \cdot \mathbb{Q} \cdot  \mathbb{M} \cdot  \kappa(v) \,, \quad  &f_2(u,v)&=2 \, \tilde \kappa(v) \cdot \mathbb{Q} \cdot  \mathbb{M} \cdot  \kappa(v)\,,\\
f_3(u,v)&=2 \, \tilde \kappa(u) \cdot \mathbb{Q} \cdot  \mathbb{M} \cdot  \tilde\kappa(v) \, , & f_4(u,v)&=2 \,  \kappa(v) \cdot \mathbb{Q} \cdot  \mathbb{M} \cdot  \kappa(v)\,,
\end{aligned}
\ee
where the $\kappa, \tilde \kappa$ are vectors with elements
\be\label{kappa_vectors}
\begin{aligned}
\kappa_{j}(u) &\equiv  -\int\limits_{0}^\infty \frac{dt}{t} \frac{J_j(2gt)(J_0(2gt)-\cos(ut)\left[e^{t/2}\right]^{(-1)^{\eta \times j}})}{e^t-1} \\
\tilde\kappa_{j}(u) &\equiv  -\int\limits_{0}^\infty \frac{dt}{t} (-1)^{j+1}\frac{J_j(2gt) \sin(u t) \left[e^{t/2}\right]^{(-1)^{\eta \times(j+1)}}}{e^t-1}
\end{aligned}
\ee
with $\eta=0$ for the NMHV case and $\eta=1$ for MHV case, and $\mathbb{Q}, \mathbb{M}$ are matrices independent of $u$. In particular, $\mathbb{Q}$ has matrix elements, $\mathbb{Q}_{ij}=\delta_{ij}(-1)^{i+1}i$, and $\mathbb{M}$ is related to another matrix $K$,
\be
\begin{aligned}
\mathbb{M} &\equiv (1+K)^{-1}=\sum_{n=0}^\infty (-K)^n\,,\\
K_{ij}&=2j(-1)^{j(i+1)} \int\limits_{0}^\infty \frac{dt}{t} \frac{J_i(2gt)J_j(2gt)}{e^t-1} \, .
\end{aligned}
\ee
The Taylor expansion of the Bessel functions $J_i$ (\ref{BesselJ_expansion}) implies that $K_{ij}$ starts at order $\mathcal{O}(g^{i+j})$. Hence even though  the vectors and matrices entering (\ref{f_definitions}) are infinite-dimensional (equivalently they are an infinite triple sum), if we wish to obtain $f_i$ up to $\mathcal{O}(g^{n})$, we can truncate them to their first $i,j=1,\ldots, n-1$ components.

The building blocks of the $f_i$ functions can also be used in order to calculate all remaining ingredients of the leading OPE contribution for the (N)MHV hexagon. Namely the anomalous dimension and momentum of the single-particle flux tube excitations can be written as
\be\label{ep_from_QM}
\gamma(u)=4 g \left( \mathbb{Q}\cdot \mathbb{M} \cdot \kappa(u) \right)_1  \, , \qquad p(u) =2u- 4 g \left( \mathbb{Q}\cdot \mathbb{M} \cdot \tilde\kappa(u) \right)_1 \, ,
\ee
and the cusp anomalous dimension as
\be
\Gamma_\text{cusp}=4g (\mathbb{Q}\cdot \mathbb{M})_{11}\,.
\ee
The expressions (\ref{ep_from_QM}), which also implicitly depend on $\eta$ by virtue of (\ref{kappa_vectors}), are equivalent to the ones used in the main text (\ref{Ep_gluon}), (\ref{Ep_scalar}), and can be derived from them.
\subsection{Harmonic Polylogarithms}\label{HPL_review}
In this appendix, we review the definition and basic properties of harmonic polylogarithms (HPL) \cite{Remiddi:1999ew}, also based on \cite{Maitre:2007kp,Ablinger:2013hcp}.

For $x\in(0,1)$ and $a_i=\{-1,0,1\}$, harmonic polylogarithms are defined as
\be\label{HPL_definition}
\begin{aligned}
H(x)&=1\,,\\
H(a_1,\ldots,a_n;x)&=\begin{cases}
\frac{1}{n!}\log^n x&\text{if }a_1=\ldots a_n=0\,,\\
\int_0^x dy f_{a_1}(t)H(a_2,\ldots,a_n;t)&\text{otherwise},
\end{cases}
\end{aligned}
\ee
where the auxiliary functions $f_a$ entering the case of the recursive definition are given by
\be
\begin{aligned}
f_{-1}(x)&=\frac{1}{1+x}\,,\\
f_{0}(x)&=\frac{1}{x}\,,\\
f_{1}(x)&=\frac{1}{1-x}\,.
\end{aligned}
\ee
The number of $a_i$ indices, or the number of integrations when the recursive definition applies, is called the weight or transcendentality of the harmonic polylogarithm. Obviously at weight $n$ we have $3^n$ distinct (but not all functionally independent) HPLs. For example for weight zero we just have $H(x)=1$, and for weight 1,
\be
\begin{aligned}
H(-1,x)&=\int_0^x f_{-1}(t)=\int_0^x \frac{1}{1+x}=\log(1+x)\,,\\
H(0,x)&=\log x\,,\\
H(1,x)&=\int_0^x f_{1}(t)=\int_0^x \frac{1}{1-x}=-\log(1-x)\,.\\
\end{aligned}
\ee

Up to weight 3, HPLs can be expressed in terms of classical polylogarithms of more general arguments \cite{Remiddi:1999ew}. However for weight 4 and higher, the latter only form a subset of the wider basis of HPL functions.

A different, more compact notation for the indices of HPLs may be obtained by replacing any string of $m-1$ subsequent zeros followed by a $\pm1$ index as follows,
\be\label{HPL_a_to_m_notation}
\overbrace{0,0,\ldots0}^{m-1},\pm1\to\pm m\,.
\ee
For example in the new notation, which is sometimes is refered to as ``m''-notation as opposed to the initial ``a''-notation,
\be
H(1,0,0,-1,0,1,0;x)=H_{1,-3,2,0}(x)\,,\quad
\ee
In the main text, we will be exclusively using the ``m''-notation.

Similarly to $Z$-sums, a product of two HPLs with the same argument can be expressed as a linear combination single HPLs. In ``a''-notation, if we denote the vector of indices $(a_1,\ldots,a_n)\equiv\mathbf{a}$, we have in particular
\be
H(\mathbf{a};x)H(\mathbf{b};x)=\sum_{\mathbf{c}=\mathbf{a}\sha \mathbf{b}}H(\mathbf{c},x)\,
\ee
where $\mathbf{a}\sha \mathbf{b}$ denotes all possible permutations of the elements of $\mathbf{a}$ and $\mathbf{b}$ combined, such that the internal order of the elements in  $\mathbf{a}$ and $\mathbf{b}$ is preserved\footnote{This operation of element mixing is called a shuffle, and the corresponding product algebra the shuffle algebra, since it is equivalent to all possible ways to riffle shuffle two decks of cards.}. For example
\be
\begin{aligned}
H(a_1,a_2;x)H(b_1,b_2;x)&=H(a_1,a_2,b_1,b_2;x)+H(a_1,b_1,a_2,b_2;x)+H(b_1,a_1,a_2,b_2;x)\\
&\,\,+H(b_1,a_1,b_2,a_2;x)+H(b_1,b_2,a_1,a_2;x)+H(a_1,b_1,b_2,a_2;x)\,,
\end{aligned}
\ee
and more generally it is not hard to see that the product of two HPLs with weights (number of indices) $w_1$ and $w_2$ will decompose into a linear combination of $(w_1+w_2)!/(w_1!w_2!)$ single HPLs.

Precisely because of the shuffle algebra, not all HPLs of a given weight will be algebraically independent, as they can be expressed in terms of other HPLs of the same weight, and products of lower weights. In any case it is always possible to construct a basis of algebraically independent HPLs, whose indices form Lyndon words \cite{Blumlein:2003gb}, although we will refrain from exploiting this property here.

Finally, an argument transformation property for HPLs with nonzero indices in the ``m''-notation, which we will be making use of, is that
\be\label{HPL_flipsign}
H_{m_1,\ldots,m_k}(-x)=(-1)^k H_{-m_1,\ldots,-m_k}(x)\,,\quad \text{for } m_k\ne0\,.
\ee
This follows simply from the definition (\ref{HPL_definition}), after we change the sign of the integration variables.

%\nocite{*}
\bibliographystyle{utphys}
\bibliography{WL_OPE}

\end{document}